\documentclass[preprint2]{aastex} 

\usepackage{natbib}
\usepackage[usenames,dvipsnames]{color}

\definecolor{dred}{rgb}{0.0,0.0,0.0}
\definecolor{dreen}{rgb}{0.0,0.0,0.0}
\definecolor{violet}{rgb}{0.0,0.0,0.0}
\definecolor{brown}{rgb}{0.0,0.0,0.0}
\definecolor{orange}{rgb}{0.0,0.0,0.0}

\definecolor{red}{rgb}{0.0,0.0,0.0}
\definecolor{blue}{rgb}{0.0,0.0,0.0}

\begin{document}

\title{Star-Forming Compact Groups (SFCGs): \\ An ultraviolet search for a
  local sample}

\author{Jonathan D. Hern\'andez-Fern\'andez\altaffilmark{1}, C. Mendes de Oliveira\altaffilmark{1}}

\altaffiltext{1}{Departamento de Astronomia, Instituto de Astronomia,
  Geof\'isica e Ci\^encias Atmosf\'ericas da Universidade de S\~ao
  Paulo, Rua do Mat\~ao 1226, Cidade Universit\'aria, 05508-090, S\~ao
  Paulo, Brazil; jonatan.fernandez@iag.usp.br}

\begin{abstract}

We present a local sample ($z$$<$0.15) of 280 Star-Forming
Compact Groups (SFCGs) of galaxies identified in the
ultraviolet Galaxy Evolution EXplorer (GALEX) All-sky
Imaging Survey (AIS). \textcolor{blue}{So far, just one
  prototypical example of SFCG, the Blue Infalling Group,
  has been studied in detail in the Local Universe.}
\textcolor{blue}{The sample of SFCGs is mainly the result of
  applying a Friends-of-Friends group finder in the space of
  celestial coordinates with a maximum linking-length of 1.5
  arcmin and choosing groups with a minimum number of four
  members of bright UV-emitting 17$<$FUV$<$20.5 sources
  (mostly galaxies) from the GALEX/AIS catalogue.} The
result from the search are 280 galaxy groups composed by
226, 39, 11 and 4 groups of four, five, six and seven bright
ultraviolet (UV) members, respectively. Only 59 of these 280
newly identified SFCGs have a previous catalogued group
counterpart. \color{blue} Group redshifts are available for
at least one member in 75\% of the SFCGs, and over 40\% of
the SFCGs have redshifts measured for two or more
galaxies. Twenty-six of the SFCGs appear to be located in
the infalling regions of clusters with known
redshift. \color{red} The SFCG sample presents a combination
of properties different from the group samples studied up to
now, such as low velocity dispersions
($\sigma_{\rm{}l-o-s}$$\sim$120\,km\,s$^{-1}$), small
crossing-times ($H_{0}$$t_{\rm{}c}$$\sim$0.05) and high
star-formation content (95\% of star-forming galaxies). This
points to the SFCGs being in an evolutionary stage distinct
from those groups selected in the optical and near-infrared
ranges. \textcolor{dreen}{Once redshifts are obtained to
  discard interlopers, SFCGs will constitute a unique sample
  of star forming compact groups.}

\end{abstract}



\section{Introduction}
\label{sec:intro} 

Over the last few years, the {\it preprocessing scenario} has been
identified as a significant mode of star-formation quenching and
morphological transformation in the context of environmental galaxy
evolution \citep{Fujita_2004,Moss_2006,Cortese_et_al_2006,
  Wilman_et_al_2009,Just_et_al_2010,Wetzel_et_al_2013,Hou_et_al_2014}. In
this scenario, environmental processes drive the evolution of galaxies
within groups before they fall into a massive cluster. In support of
this scenario, \citet{McGee_et_al_2009} found that simulated clusters
up to $z$=1.5 have had a significant fraction of their galaxies
accreted through galaxy groups. For instance,
10$^{14.5}$$h^{-1}$~M$_{\odot}$ mass clusters at $z$=0 have had
$\sim$40 per cent of their galaxies with stellar masses
$M_{\star}$$>$10$^{9}$$h^{-1}$~M$_{\odot}$ accreted through haloes
with masses greater than 10$^{13}$$h^{-1}$~M$_{\odot}$. At higher
redshifts, fewer galaxies are accreted through massive haloes. Only
$\sim$25 per cent of galaxies have been accreted through
10$^{13}$~M$_{\odot}$ groups into 10$^{14.5}$$h^{-1}$~M$_{\odot}$
clusters at $z$=1.5.

\begin{figure}[!ht]
\includegraphics[width=1.00\hsize]{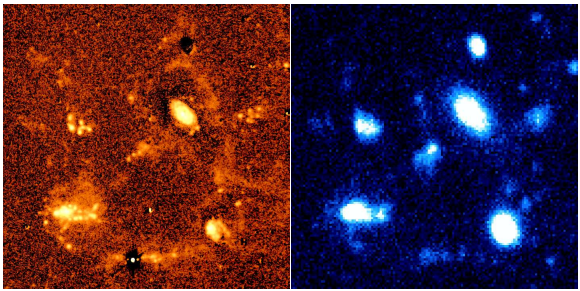}
\caption{\bf Two views of the Blue Infalling Group (BIG). Left panel:
  H$\alpha$ net-flux image of the BIG taken from GOLD~Mine database
  (http://goldmine.mib.infn.it). Right panel: GALEX NUV image of the
  BIG. The size of the image is 3.25$\times$3.25 arcmin$^{2}$,
  corresponding to 108 kpc.}
\label{fig:Ha_UV_BIG}
\end{figure}

The unique case in the local Universe considered as an ongoing example
of the preprocessing scenario is the {\it Blue Infalling Group} (BIG);
the region with the highest density of star forming systems ever
observed in the Local Universe \citep{Cortese_et_al_2006}. Only this
case of an infalling group with bursts of star formation induced by
ram-pressure has been observed in H$\alpha$ in the Local Universe and
the group shows a wealth of structures hosting current star formation
activity such as tidal streams or clustered knots, as it can be seen
in figure \ref{fig:Ha_UV_BIG}. Similar examples of star-forming
infalling groups in the Local Universe are expected to be rare since
the current merger rate in clusters is considerably lower than in the
past \citep[e.g.][]{Gottlober_et_al_2001}. In figure
\ref{fig:Ha_UV_BIG} we show the comparison between the H$\alpha$
net-flux (left panel) and the Galaxy Evolution EXplorer (GALEX) NUV
image (right panel) of the BIG. As it can be easily seen in this
figure, even in the cases where there are just faint clustered knots
of H$\alpha$ emission in the optical image, bright ultraviolet (UV)
counterparts can be found in the GALEX image. Consequently, GALEX
catalogues provide an excellent chance to find analogues of BIG.

The {\it double-cycle scenario} recently proposed by
\citet{Dressler_et_al_2013} presents some tension with the view of BIG
as an ongoing event of group preprocessing. The authors, studying a
sample of intermediate redshift clusters, point that poststarburst
spectra trace accretions in passive galaxies, while a starburst
spectra signal minor mergers in star-forming galaxies. In this
picture, spectroscopically identified post-starbursts constitute a
minority of all recently terminated starbursts, largely ruling out the
typical starburst as a quenching event in all but the densest
environments. A sample of Star-Forming Compact Groups (SFCGs)
analogues of the BIG, can be an excellent chance to assess the
relative predominance of starbursts and post-starbursts in the process
of star-formation quenching and morphological transformation within
infalling groups.

Differently from galaxies in clusters, {\it galaxies in groups}
represent a large contribution of the giant galaxy population in the
nearby universe \citep{Cox_2000,Eke_et_al_2005}. \textcolor{blue}{More
  specifically, for a local volume-limited sample of luminous
  $M_{r}$$<$$-$18 galaxies from \citet{Nurmi_et_al_2013}, 40\% of
  galaxies are considered isolated, 16\% are galaxies in pairs, 35\%
  of the galaxy population is in groups of 3$\le$$n_{\rm{}gal}$$<$30
  members and just 9\% belong to galaxy clusters of 30 members or
  more. \citet{Nurmi_et_al_2013} apply a Friends-of-Friends (FoF)
  algorithm with a variable linking-length depending on redshift and
  calibrated with a local sample of real galaxy groups.} Further,
detailed studies of galaxy interactions and environmental processes in
their most frequent environment in the very moment when they occur are
indispensable to adequately describe galaxy evolution. In this
respect, H$\alpha$ imaging allows us to identify where within the
group galaxies they are forming stars normally and where they have had
their star formation enhanced or depressed by interactions with the
other group members and other environmental processes i.e. gas
stripping, starvation \citep{Koopmann&Kenney_2004}. Interestingly,
SFCGs are ideal targets where H$\alpha$ emission maps are more easily
obtained, particularly for those instruments with relatively small
fields of view \citep[e.g.~BTFi,][]{MdO_BTFI}.

As an additional product of strong {\it galaxy interactions}, stellar
material may be ejected in the surroundings. In the situation when
this material is dynamically bound, it may be the progenitor of a
population of the so-called tidal dwarf galaxies
\citep{Duc_et_al_2000,Iglesias-Paramo&Vilchez_2001} and/or HII
extragalactic clumps \citep{MdO_et_al_2004,
  Boquien_et_al_2010,Urrutia-Viscarra_et_al_2014}. Otherwise, it would
contribute to the intra-group(-cluster) light. In this respect,
\citet{Domainko_et_al_2006} show that ram-pressure stripping can
account for $\sim$10\% of the overall observed level of metal
enrichment in the intracluster medium (ICM) within the virial
region. A detailed analysis of the frequency and properties of the
population of dwarf galaxies, HII extragalactic regions and intragroup
light in the sample of SFCGs would provide insights about the
contribution of galaxy interactions to group/cluster components.

It is broadly understood that galaxy groups go through an {\it
  evolutionary sequence} starting in poor groups as the Local Group
and evolving towards those galaxy groups dominated by a very massive
elliptical galaxy called fossil groups
\citep[after][]{Ponman_et_al_1994}, or less massive versions of them
\citep{Jones_et_al_2003,DOnghia_et_al_2005,
  Zabludoff_2007,Grebel_2007,Gallagher_et_al_2010},
\textcolor{blue}{or yet massive elliptical galaxies in the field
  \citep{Mamon_1986,Williams&Rood_1987}.} In this scenario, {\it
  compact groups} represent the evolutionary link between these
apparently unrelated extragalactic objects.

Such as it can be seen in literature, a detailed study of a sample of
compact groups of star-forming galaxies can shed light in a broad
ensemble of aspects of galaxy evolution in groups. \color{red} Section
\ref{sec:gal_samp} describes the selection of a sample of this type,
starting from catalogues of pure UV sources and also presents the
final sample including observational information and basic
properties. In Section \ref{sec:chcomp_gsamp}, several aspects of the
SFCG sample are analyzed and compared with three other prototypical
compact group samples: the optically-selected Hickson Compact Group
\citep[HCG,][]{Hickson_1982,Hickson_et_al_1992} sample, the
near-infrared-selected and velocity-filtered 2MASS Compact Group
\citep[2MCG,][]{Diaz-Gimenez_et_al_2012} sample \color{dreen} and the
catalogue~A of compact groups selected in the Sloan Digital Sky Survey
by \citet[SCGA,][]{McConnachie_et_al_2009}. \color{black} Section
\ref{sec:x-matching} places the SFCG sample in the context of the
Local Universe providing additional information about their galaxy
members, cluster and group counterparts and also, the close
surroundings of its groups.


\section{Selection of the SFCG sample}
\label{sec:gal_samp}

Our approach is focused on searching for compact groups composed or
dominated by star-forming galaxies in the Local Universe.

\subsection{An all-sky sample of bright ultraviolet-emitting sources}
\label{ssec:uvsample}

We choose the All-sky Imaging Survey (AIS) from GALEX as the most
suitable database to extract large samples of star-forming galaxies in
the Local Universe. The GALEX/AIS is the largest effective sky area,
more than 22000 square degrees, observed in UV in a homogeneous way
down to a typical depth of FUV$\sim$20 AB mag and NUV$\sim$21 AB mag
\citep{Bianchi_et_al_2014}.

We select those UV sources with an apparent FUV magnitude of
17$<$FUV$<$20.5, because the FUV band is tracing the emission of stars
with shorter lifetimes in comparison with the NUV
\citep{Martin_et_al_2005,Haines_et_al_2008}. We note that the
UV-bright galaxies in BIG are approximately in this magnitude
range. This small magnitude interval is chosen to increase the
probability of selecting galaxies that are physically associated with
each other (by not being at greatly differing distances), and with a
young stellar component which is similar so that all group members
would be significant participants in the total star formation of the
group. Given that the redshift distribution of GALEX sources is mostly
restricted to $z$$\lesssim$0.2 \citep{Wyder_et_al_2005}, this further
biases the selection towards galaxies with a high level of star
formation activity in the Local Universe.

\textcolor{blue}{We stress that the selection using the GALEX FUV band
  allows us to also include in our sample the low-mass star-forming
  galaxies, without a bright counterpart in the optical bands, and
  therefore, which rarely included in searches of groups performed in
  the optical and near-infrared \citep[i.e.][]{
    Hickson_1982,Hickson_et_al_1992,McConnachie_et_al_2009,
    Diaz-Gimenez_et_al_2012}. The searches of galaxy groups extracted
  from optical and near-infrared surveys favour galaxy groups
  constituted by the massive examples of the galaxy population i.e. a
  {\it stellar-mass biased search}, while a search of groups of
  UV-bright galaxies finds those galaxies which encompass the bulk of
  the current star formation of the galaxy population, i.e.  a {\it
    star-formation biased search}.  A test of an ultraviolet search
  over the galaxy members of the HCG catalogue is included in
  Subsection \ref{ssec:uvhick} and confirms this statement. In spite
  of that, one has to keep in mind that the SFCGs can indeed contain
  passive galaxies as close companions of those star-forming galaxies
  with a high UV emission.}

In addition, the following UV colour constraint
$-$1.50$\le$(FUV$-$NUV)$_{\rm{}d}$$\le$2.75 is included in the
source selection. (FUV$-$NUV)$_{\rm{}d}$ corresponds to the GALEX
UV colour corrected from Galactic extinction, following the Cardelli
extinction law \citep{Cardelli_et_al_1989} and a value of the ratio of
total to selective absorption in the V\,band of $R_{\rm{}V}$=3.1. This
has been shown to be very useful in rejecting `blue artifacts',
commonly found throughout GALEX fields, particularly in the
neighbourhood of saturated stars and also over galactic regions with a
diffuse UV emission, which do not correspond to real sources. Indeed,
this constraint could reject some real UV galaxies but we prefer to
include this condition for the sake of purity of the input UV sample
and the efficiency of the search.

\color{red} Furthermore, we take advantage of the GALEX flag {\tt
  nuv\_artifact} which is a photometric-quality estimator. {\tt
  nuv\_artifact} consists in a bitwise-encoded unsigned integer, each
of which is the bitwise logical 'OR' of some or none of a set of nine
possible situations (see
http://galex.stsci.edu/\allowbreak{}GR6/?page=ddfaq\#6 for a detailed
description of each situation). Criteria about the source brightness
for flagging one source are generous, meaning that false-positives are
common, especially for `edge' flags i.e. {\tt\,nuv\_artifact}=1. {\tt
  nuv\_artifact} increases its value as a larger number of situations
affects the UV source. We therefore consider that selecting sources
with {\tt\,nuv\_artifact$\le$1} is sufficient for a complete and
sufficiently reliable source extraction.

\color{black} Given that we search for compact groups from a sample
without information about the nature of the UV sources, we do not know
a priori whether the UV source will be a star or a galaxy. Because of
that, \color{blue} we try to minimize the effects of contamination by
stars by restricting our search to sources beyond 15 degrees of the
Galactic Plane i.e. $\mid$b$\mid$$\ge$15\,deg.

\color{black} We download a set of photometric parameters for a sample
with the previous constraints running the following SQL code in the
CASJobs service of GALEX webpage ({\it
  http://galex.stsci.edu/casjobs}):

{\tt SELECT p.ra, p.dec, p.fuv\_mag, p.nuv\_mag, \\
p.e\_bv, p.glon, p.glat, q.photoExtractID, \\ 
p.objid, q.mpstype \\
FROM PHOTOOBJALL as p, PHOTOEXTRACT as q \\
WHERE (p.fuv\_mag BETWEEN 17.0 and 20.5) \\
AND (((fuv\_mag$-$(E\_bv$\ast$8.376)) \\
$_{}$~~~~ $-$(nuv\_mag$-$(E\_bv$\ast$8.741))) \\ 
BETWEEN $-$1.50 AND +2.75 ) \\
AND ((p.glat$<$$-$15.0) OR (p.glat$>$+15.0)) \\ 
AND (p.nuv\_artifact$<$=1) \\
AND (q.mpstype=$'$AIS$'$) \\
AND (p.photoExtractID=q.photoExtractID)}
 
The result from this SQL query is a catalogue containing 925428
sources.

Due to the GALEX survey strategy, sky regions at the outermost parts
of the fields and the UV sources in these regions are systematically
observed by neighboring fields i.e. there is significant
overlapping. \textcolor{blue}{We identify duplicate objects by
  cross-matching the GALEX sample with itself, using a 6 arcsec search
  radius, slightly larger than 4.2 and 5.3 arcsec angular resolution
  of the GALEX FUV and NUV images, respectively
  \citep{Morrissey_et_al_2007}.}

\subsection{Grouping UV sources with the Friends-of-Friends Algorithm}

\textcolor{blue}{The search for compact groups is done applying a
  Friends-of-Friends (FoF) algorithm to the previous sample of bright
  UV sources in the space of celestial coordinates imposing a maximum
  linking-length.  This approach was originally applied by
  \citet{Turner&Gott_1976} to a sample of galaxies from the {\it
    Catalogue of galaxies and of clusters of galaxies} of
  \citet{Zwicky_et_al_1961,Zwicky_et_al_1968}. The upper limit of the
  linking-length is set to 1.5 arcmin, which corresponds to a
  projected distance of $\sim$88 kpc at $z$=0.05 in the cosmological
  model used in this work and characterized by
  $H_{0}$=70\,km\,s$^{-1}$Mpc$^{-1}$, $\Omega_{\rm{}m}$=0.3 and
  $\Omega_{\rm{}\Lambda}$=0.7.} This criterion is included to accept
only those groups that are sufficiently compact and also, easily
observable with relatively small field instrumentation
\citep[e.g.][]{MdO_BTFI}. The IDL code of the FoF
algorithm\footnote{http://spectro.princeton.edu/idlutils\_doc.html\#SPHEREGROUP}
(Blanton\,2001) selects a set of groups whose elements are linked by
sky distances less or equal to the linking-length, labeling each group
by their number of elements. This code belongs to the IDLUTILS library
from the Princeton/MIT Sloan Spectroscopy Home Page
(http://spectro.princeton.edu/idlutils\_doc.html). \textcolor{blue}{The
  output group catalogue contains 1447 compact groups of UV bright
  sources including both galaxy and stellar objects. We have found
  that about 960 of these groups surround the Large and the Small
  Magellanic Clouds and the bulk of them are projected groups of open
  star clusters or parts of these two galaxies.}

\subsection{Cross-matching with galaxy catalogues}
\label{ssec:+match}

Knowing that the search of groups of UV sources is done without {\it a
  priori} information about the object type and the redshift of the UV
source, we crossmatch the output sample of groups of UV sources with
the galaxy catalogues compiled by NASA/IPAC Extragalactic Database
(NED) applying a matching radius of 6\,arcsec in order to retrieve the
object type and redshift of each UV source. The crossmatching is
provided by NED through the service ``Build Data Table from Input
List" in its option ``Near Name/Position Cross-Matching":
(http://ned.ipac.caltech.edu/forms/nnd.html). In the few cases that
NED provides more than one counterpart inside the matching circle, we
choose an unique source following this hierarchy in priority (1) the
galaxy with redshift counterpart (2) the source classified as
``galaxy" (3) the nearest source. 

\subsection{Selection of galaxy group candidates and SFCG catalogue}
\label{ssec:gselec}

We only include in the final catalogue of SFCGs those groups with four
or more UV sources ($n_{\rm{}UV}$$\ge$4) which respect at least one of
the two following criteria:

\begin{itemize}

\item groups with at least three UV sources compiled as ``galaxy" by
  NED $n_{\rm{}gal}$$\ge$3 (i.e. three galaxies) or,

\item groups with at least two galaxies with redshift counterpart
  within a redshift interval of
  \textcolor{red}{$\Delta$$z$/(1+$z$)=0.004} (i.e. two accordant
  redshifts)

\end{itemize}

Visually checking the close neighborhood of the groups of UV-emitting
sources in GALEX and optical images, we know that these groups can
include galaxies without UV emission (i.e. red galaxies) and/or UV
galaxies that are so optically faint that they are missed by
NED. Knowing this fact, we want to be generous in these first
constraints, given that the actual number of galaxy members in each
group should be greater than its number of UV sources ($n_{\rm{}UV}$)
or the number of UV sources compiled as galaxies by NED
($n_{\rm{}gal}$).

The previous constraints give an output of 280 compact groups of
UV-emitting galaxies with the following \textcolor{red}{multiplicity
  distribution $N$($n$)}: 226, 39, 11 and 4 groups with four, five,
six and seven UV members, respectively. \color{blue} This multiplicity
distribution follows very accurately a power law as
$N$($n$)$\thicksim$$n^{-\alpha}$ with
$\alpha_{\rm{}UV}$=7.54$\pm$0.49. As a reference the HCGs follows a
multiplicity distribution also well fitted by a power law with
$\alpha_{\rm{}HCG}$=5.25$\pm$0.62. Therefore, the multiplicity
distribution of the SFCG sample is steeper than the multiplicity
distribution of the HCG sample. The latter sample is, in addition,
poor in UV-bright star-forming galaxies as it is shown in Subsection
\ref{ssec:uvhick}. This fact is in general agreement with the weaker
clustering of blue galaxies in comparison with red galaxies
\citep[e.g.][]{Zehavi_et_al_2011}.

\color{black} Celestial coordinates, group redshift ($z_{\rm{}UV}$)
when available and information in other group properties are shown in
Table \ref{tab:CS}, where SFCGs are ordered by increasing UV richness
$n_{\rm{}UV}$ and increasing right ascension. \color{red} Table
\ref{tab:CgalS} compiles celestial coordinates, NED redshifts when
available and GALEX magnitudes for the group UV-bright galaxy members
ordered by increasing group ID and increasing right
ascension. \color{dreen} The values for $z_{\rm{}UV}$ correspond to
the median redshift of the $n_{z}$ UV galaxies with known redshift.
In Table \ref{tab:CS}, $m_{z}$ is the number of galaxies with known
redshifts within a velocity interval of
$\mid$$\Delta$v$_{\rm{}l-o-s}$$\mid$$\le$ 10$^{3}$\,km\,s$^{-1}$,
where $\Delta$v$_{\rm{}l-o-s}$ is the l-o-s velocity difference with
respect to the group redshift i.e.:

\begin{equation}
\Delta{\rm{}v}_{\rm{}l-o-s}={\rm{}c}~\left(\frac{z_{\rm{}memb}-z_{\rm{}UV}}{1+z_{\rm{}UV}}\right).
\label{eq:dvlos}
\end{equation}

If there is no known redshift associated to the UV galaxies of a
specific SFCG, i.e. $n_{\rm{}z}$=0, we label the redshift value with
zero in Table \ref{tab:CS}, $z_{\rm UV}$$\equiv$0. \color{black} This
approach yields 212 group redshifts ($\approx$76\%) of the whole
sample), among which \color{brown} 97 SFCGs with $m_{z}$=1
($\approx$35\%), 74 SFCGs with $m_{z}$=2 ($\approx$26\%), 36 SFCGs
with $m_{z}$=3 ($\approx$13\%) and 5 SFCGs with $m_{z}$=4
($\approx$2\%). There are no SFCGs with more than
$m_{z}$=4. \color{blue} In addition, we provide the angular radius
$r_{\theta}$ of each SFCG, computed as the square-root of the
quadratic average of angular distances $r_{i}$ of galaxy members to
the group center i.e.:

\begin{equation}
r^{2}_{\theta}=\left(~\sum{r_{i}^{2}}~\right)~/~(n_{\rm{}UV}-1).
\label{eq:rtheta}
\end{equation}

\color{black} We also provide a collection of stamps for the whole
SFCG sample in the optical and ultraviolet spectral ranges. The
optical images come from the R\,band of the POSS2/UKSTU surveys and
the ultraviolet images correspond to the NUV\,band from
GALEX. \color{dreen} The images are fully public and added as
additional on-line only material in Figure A1 of Appendix.

\onecolumn

\begin{table}[!ht]
\color{black}
\caption{\bf SFCG catalogue}
\begin{center}
\resizebox{!}{0.40\vsize}{{\bf 
\begin{tabular}{r rr l cccc c rr}
\hline
      ID    & RA$_{\rm{}g}$    & Dec$_{\rm{}g}$     & z$_{\rm{}UV}$  & n$_{\rm{}UV}$ & n$_{\rm{}gal}$  & n$_{z}$  & m$_{z}$   & $r_{\rm{}\theta}$  &  $R_{ij}$  &  $\sigma_{\rm{}l-o-s}$ \\
            & [h:m:s]        &  [d:m:s]         &              &             &              &          &          &  [arcmin]        &   [kpc]   &   [km\,s$^{-1}$]      \\
     (1)    &      (2)       &      (3)         &       (4)    &  (5)        & (6)          &     (7)  &  (8)     &   (9)            &    (10)   &  (11)                \\
\hline                                                                                                  
       1 &      0:00:23 &    -22:35:14 &   0        &        4 &        4 &        0 &        0 &     0.913   &       -1~~~ &       -1~~~ \\
       2 &      0:01:23 &     13:06:32 &   0.017899 &        4 &        4 &        3 &        3 &     1.611   &       32.94 &      193.85 \\
       3 &      0:11:22 &    -53:57:20 &   0.041719 &        4 &        3 &        1 &        1 &     1.249   &       -1~~~ &       -1~~~ \\
       4 &      0:13:19 &    -41:34:47 &   0        &        4 &        3 &        0 &        0 &     1.027   &       -1~~~ &       -1~~~ \\
       5 &      0:14:09 &    -22:33:51 &   0        &        4 &        4 &        0 &        0 &     1.149   &       -1~~~ &       -1~~~ \\
       6 &      0:17:46 &    -69:08:55 &   0        &        4 &        3 &        0 &        0 &     1.336   &       -1~~~ &       -1~~~ \\
       7 &      0:18:29 &    -42:07:39 &   0.093258 &        4 &        3 &        1 &        1 &     0.891   &       -1~~~ &       -1~~~ \\
       8 &      0:18:39 &    -35:57:01 &   0.012953 &        4 &        4 &        2 &        2 &     0.888   &       -1~~~ &       -1~~~ \\
       9 &      0:21:29 &     38:05:07 &   0.035708 &        4 &        3 &        2 &        2 &     1.161   &       -1~~~ &       -1~~~ \\
      10 &      0:22:41 &    -20:34:38 &   0.051799 &        4 &        3 &        1 &        1 &     1.041   &       -1~~~ &       -1~~~ \\
      11 &      0:31:14 &    -25:29:44 &   0.092950 &        4 &        4 &        2 &        2 &     1.091   &       -1~~~ &       -1~~~ \\
      12 &      0:34:30 &     39:35:15 &   0.019458 &        4 &        2 &        2 &        2 &     1.634   &       -1~~~ &       -1~~~ \\
      13 &      0:37:08 &    -63:26:26 &   0.041095 &        4 &        3 &        1 &        1 &     0.984   &       -1~~~ &       -1~~~ \\
      14 &      0:37:42 &     -9:03:24 &   0.076528 &        4 &        4 &        3 &        3 &     0.761   &       74.83 &      228.04 \\
      15 &      0:37:56 &    -25:04:18 &   0.063567 &        4 &        3 &        3 &        2 &     1.152   &       -1~~~ &       -1~~~ \\
      16 &      0:48:02 &    -63:09:10 &   0.086353 &        4 &        4 &        2 &        1 &     1.087   &       -1~~~ &       -1~~~ \\
      17 &      0:48:58 &    -45:46:13 &   0        &        4 &        4 &        0 &        0 &     0.809   &       -1~~~ &       -1~~~ \\
      18 &      0:49:11 &    -18:17:43 &   0.159397 &        4 &        3 &        1 &        1 &     1.461   &       -1~~~ &       -1~~~ \\
      19 &      0:51:27 &    -23:12:56 &   0        &        4 &        3 &        0 &        0 &     0.658   &       -1~~~ &       -1~~~ \\
      20 &      0:52:11 &     -1:07:02 &   0.041300 &        4 &        4 &        3 &        2 &     0.596   &       -1~~~ &       -1~~~ \\
     ... &          ... &          ... &        ... &      ... &      ... &      ... &      ... &       ...                               \\
     261 &     17:49:58 &     68:24:19 &   0.051225 &        5 &        4 &        3 &        3 &     1.190   &       53.04 &      219.04 \\
     262 &     21:27:56 &    -46:38:34 &   0.046689 &        5 &        5 &        1 &        1 &     0.979   &       -1~~~ &       -1~~~ \\
     263 &     21:35:23 &    -49:56:05 &   0.054924 &        5 &        3 &        2 &        2 &     0.870   &       -1~~~ &       -1~~~ \\
     264 &     23:16:44 &    -41:39:20 &   0        &        5 &        4 &        0 &        0 &     1.304   &       -1~~~ &       -1~~~ \\
     265 &     23:56:08 &    -17:14:21 &   0        &        5 &        5 &        0 &        0 &     0.990   &       -1~~~ &       -1~~~ \\
     266 &      0:25:12 &    -60:37:48 &   0.031292 &        6 &        6 &        1 &        1 &     1.141   &       -1~~~ &       -1~~~ \\
     267 &      1:39:14 &    -74:25:23 &   0        &        6 &        3 &        0 &        0 &     1.023   &       -1~~~ &       -1~~~ \\
     268 &      2:34:18 &    -73:43:12 &   0        &        6 &        3 &        0 &        0 &     1.339   &       -1~~~ &       -1~~~ \\
     269 &      3:21:01 &    -33:54:17 &   0.070421 &        6 &        5 &        4 &        3 &     1.144   &      179.78 &      224.71 \\
     270 &      4:06:11 &    -36:55:48 &   0        &        6 &        4 &        0 &        0 &     0.995   &       -1~~~ &       -1~~~ \\
     271 &      4:49:33 &    -73:04:09 &   0.027543 &        6 &        3 &        2 &        2 &     1.263   &       -1~~~ &       -1~~~ \\
     272 &      8:20:05 &     21:04:17 &   0.017462 &        6 &        4 &        3 &        1 &     1.583   &       -1~~~ &       -1~~~ \\
     273 &     11:14:32 &      0:51:08 &   0.069070 &        6 &        6 &        5 &        4 &     1.519   &      106.41 &       57.16 \\
     274 &     12:21:14 &     49:27:34 &   0.045355 &        6 &        5 &        4 &        2 &     1.354   &       -1~~~ &       -1~~~ \\
     275 &     15:21:59 &      3:33:36 &   0.084655 &        6 &        5 &        4 &        4 &     0.701   &       72.12 &      268.79 \\
     276 &     21:55:29 &    -20:51:46 &   0.066913 &        6 &        5 &        1 &        1 &     1.335   &       -1~~~ &       -1~~~ \\
     277 &      2:05:03 &    -74:51:50 &   0        &        7 &        3 &        0 &        0 &     1.597   &       -1~~~ &       -1~~~ \\
     278 &      7:41:56 &     16:49:33 &   0        &        7 &        4 &        0 &        0 &     1.103   &       -1~~~ &       -1~~~ \\
     279 &     13:07:17 &     13:38:57 &   0.063226 &        7 &        6 &        5 &        1 &     1.704   &       -1~~~ &       -1~~~ \\
     280 &     23:16:45 &      9:48:57 &   0        &        7 &        3 &        0 &        0 &     1.114   &       -1~~~ &       -1~~~ \\
\hline
\end{tabular} } }
\end{center}
{\bf \color{dreen} (1) ID number of the SFCG ordered by increasing UV richness
  (column 5) and increasing right ascension (column 2), 
  (2) and (3) R.A. [h:m:s] and Dec. [d:m:s] of group center, 
  (4) group redshift,
  (5) number of UV members, 
  (6) $n_{\rm{}UV}$ compiled as `galaxy' by NED, 
  (7) $n_{\rm{}UV}$ with a redshift counterpart, 
  (8) $n_{\rm{}UV}$ with a known l-o-s velocity within 2000\,km\,s$^{-1}$ centered at the group redshift,
  (9) group angular radius in arcmin, \color{brown} 
  (10) median inter-galaxy physical separation in kpc and
  (11) l-o-s velocity dispersion derived by the gapper method
  \citep{Beers_et_al_1990}. For $m_{\rm{}z}$=3, the relative
  uncertainty in $\sigma_{\rm{}l-o-s}$ corresponds to 50\% and for
  $m_{z}$=4 corresponds to 41\% following
  ($\sigma$/$\delta$$\sigma$)=$\sqrt{2(m_{z}-1)}$
  \citep[e.g.][]{Proctor_et_al_2011}.}
\label{tab:CS}
\end{table}
\newpage

\begin{table}[!ht]
\color{red}
\caption{\bf SFCG galaxy catalogue}
\begin{center}
{\bf 
\begin{tabular}{cc rr l r cc}
\hline
 ID  &  ID$_{\rm{}memb}$  &  RA$_{\rm{}memb}$  &   Dec$_{\rm{}memb}$  & $z_{\rm{}memb}$  & $g_{\rm{}memb}$ & FUV$_{\rm{}memb}$ & NUV$_{\rm{}memb}$ \\
     &                 &   [deg[          &   [deg]            &               &               &  [AB mag] &   [AB mag]           \\
 (1) &  (2)            &    (3)           &   (4)              &   (5)         &  (6)          &   (7)     &     (8)              \\
\hline
    1   &     1   &       0:00:18.98   &    -22:34:33.2      &      0          &        0   &  20.2714   &  19.9779  \\ 
    1   &     2   &       0:00:21.87   &    -22:35:16.4      &      0          &        0   &  20.2043   &  19.9377  \\ 
    1   &     3   &       0:00:24.05   &    -22:35:19.7      &      0          &        0   &  19.3392   &  19.2220  \\ 
    1   &     4   &       0:00:26.82   &    -22:35:44.9      &      0          &        0   &  20.0727   &  19.8536  \\ 
    2   &     1   &       0:01:15.17   &     13:06:48.3      &      0.018646   &        1   &  18.5922   &  18.1860  \\ 
    2   &     2   &       0:01:20.07   &     13:06:40.3      &      0.017899   &        1   &  16.7561   &  16.2019  \\ 
    2   &     3   &       0:01:26.22   &     13:06:45.1      &      0.017532   &        1   &  17.1312   &  16.6332  \\ 
    2   &     4   &       0:01:29.52   &     13:05:54.9      &      0          &        0   &  19.2810   &  19.2182  \\ 
    3   &     1   &       0:11:20.38   &    -53:57:21.6      &      0          &       -1   &  20.2101   &  20.5555  \\ 
    3   &     2   &       0:11:21.11   &    -53:58:51.2      &      0          &        0   &  20.0492   &  19.6655  \\ 
    3   &     3   &       0:11:22.08   &    -53:57:13.3      &      0          &        0   &  19.9107   &  19.7047  \\ 
    3   &     4   &       0:11:24.55   &    -53:55:52.3      &      0.041719   &        0   &  17.7296   &  17.3873  \\ 
    4   &     1   &       0:13:15.43   &    -41:34:16.3      &      0          &        0   &  20.2139   &  19.7874  \\ 
    4   &     2   &       0:13:16.15   &    -41:35:44.2      &      0          &        0   &  20.2478   &  19.4746  \\ 
    4   &     3   &       0:13:18.35   &    -41:34:23.5      &      0          &       -1   &  18.2981   &  18.3077  \\ 
    4   &     4   &       0:13:24.64   &    -41:34:42.6      &      0          &        0   &  19.8293   &  19.2635  \\ 
    5   &     1   &       0:14:04.24   &    -22:33:09.7      &      0          &        0   &  20.2563   &  20.0467  \\ 
    5   &     2   &       0:14:07.62   &    -22:33:47.2      &      0          &        0   &  19.9924   &  19.2451  \\ 
    5   &     3   &       0:14:09.35   &    -22:34:04.8      &      0          &        0   &  20.1528   &  19.8885  \\ 
    5   &     4   &       0:14:14.88   &    -22:34:22.1      &      0          &        0   &  19.9073   &  19.6148  \\ 
  ...   &   ...   &          ...       &        ...          &     ...         &      ...   &    ...     &    ...    \\ 
  278   &     2   &       7:41:53.17   &     16:49:43.9      &      0          &        0   &  19.1374   &  18.9651  \\ 
  278   &     3   &       7:41:53.85   &     16:49:57.6      &      0          &        0   &  19.9631   &  19.5871  \\ 
  278   &     4   &       7:41:55.01   &     16:49:55.1      &      0          &       -1   &  20.0733   &  19.3963  \\ 
  278   &     5   &       7:41:57.17   &     16:49:30.1      &      0          &        0   &  19.8613   &  21.1990  \\ 
  278   &     6   &       7:41:59.39   &     16:49:31.6      &      0          &        0   &  18.7801   &  18.2003  \\ 
  278   &     7   &       7:42:02.43   &     16:48:30.7      &      0          &       -1   &  17.3001   &  17.6996  \\ 
  279   &     1   &      13:07:10.80   &     13:40:50.6      &      0          &       -1   &  19.9962   &  19.8683  \\ 
  279   &     2   &      13:07:11.70   &     13:39:33.1      &      0.027162   &        0   &  18.7777   &  18.2494  \\ 
  279   &     3   &      13:07:12.74   &     13:38:48.5      &      0.100102   &        0   &  20.2559   &  19.7801  \\ 
  279   &     4   &      13:07:17.56   &     13:38:47.7      &      0.026784   &        0   &  18.6093   &  18.0461  \\ 
  279   &     5   &      13:07:19.04   &     13:38:24.2      &      0          &        0   &  17.8783   &  17.3566  \\ 
  279   &     6   &      13:07:21.07   &     13:37:43.7      &      0.099919   &        0   &  20.0560   &  19.1825  \\ 
  279   &     7   &      13:07:26.39   &     13:38:29.6      &      0.063226   &        1   &  19.2524   &  18.8211  \\ 
  280   &     1   &      23:16:42.17   &      9:50:15.5      &      0          &       -1   &  19.6095   &  19.3379  \\ 
  280   &     2   &      23:16:42.63   &      9:47:32.3      &      0          &        0   &  18.3754   &  17.8370  \\ 
  280   &     3   &      23:16:43.20   &      9:48:58.2      &      0          &       -1   &  19.2525   &  18.8421  \\ 
  280   &     4   &      23:16:43.27   &      9:49:44.8      &      0          &       -1   &  19.2900   &  18.8361  \\ 
  280   &     5   &      23:16:47.03   &      9:48:45.5      &      0          &       -1   &  19.3780   &  19.0168  \\ 
  280   &     6   &      23:16:47.67   &      9:48:17.6      &      0          &        0   &  18.2314   &  17.5936  \\ 
  280   &     7   &      23:16:48.49   &      9:49:02.0      &      0          &        0   &  19.2150   &  18.9493  \\ 
\hline
\end{tabular} } 
\end{center}
\color{dreen}
{\bf (1) ID number of the SFCG, 
     (2) ID number of the SFCG galaxy ordered by increasing right ascension (column 3), 
     (3) and (4) R.A. [h:m:s] and Dec. [d:m:s] of UV-bright galaxy member, 
     (5) galaxy redshift, 
     (6) group membership code (1: velocity-accordant galaxy member, 0: non-member galaxy or unknown redshift, -1: non-NED galaxy),
     (7) and (8) FUV and NUV (AB) magnitudes corrected for Galactic extinction.}
\label{tab:CgalS}
\end{table}
\twocolumn 
\newpage

\section{Characterization of the SFCG sample and comparison with other groups samples}
\label{sec:chcomp_gsamp}

\subsection{Redshift and physical size distribution and the comparison with other group samples}
\label{ssec:zphy}

\begin{figure}[!ht]
\centering
\resizebox{1.00\hsize}{!}{\includegraphics{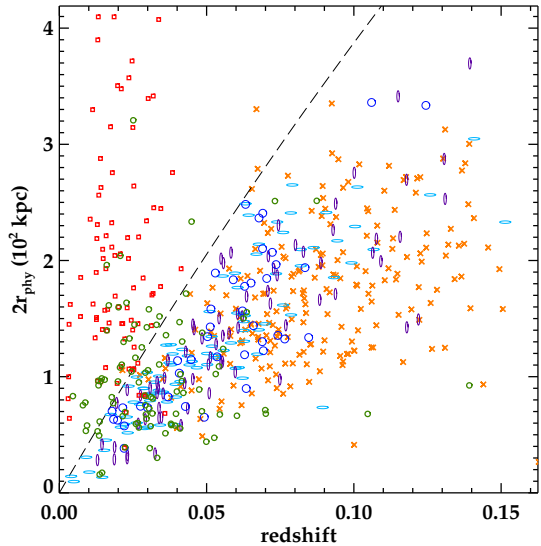}}
\caption{\bf \color{dreen} Physical diameter 2$r_{\rm{}phy}$ versus
  redshift. Cyan horizontal and violet vertical ellipses correspond to
  those SFCGs with $m_{z}$=1 and $m_{z}$=2, respectively. Those SFCGs
  with $m_{z}$$\ge$3 are represented by large blue circles. Small
  green circles, red squares and orange crosses correspond to the HCG,
  2MCG and SCGA sample, respectively. Dashed line corresponds to an
  angular diameter of 2$r_{\theta}$=3.5 arcmin.}
\label{fig:phys_z}
\end{figure}

\color{dreen} Figure \ref{fig:phys_z} shows the redshift and physical
diameter distribution of the subsets of SFCGs, the HCGs and the SCGAs
with $m_{z}$$\ge$3 i.e. three or more galaxy members within a l-o-s
velocity
$\mid$$\Delta$v$_{\rm{}l-o-s}$$\mid$$\le$10$^{3}$\,km\,s$^{-1}$
centered at the group redshift. The 2MCG sample has $m_{z}$$\ge$4 by
construction. Figure \ref{fig:phys_z} also shows the SFCG subsets with
$m_{z}$=1 (horizontal cyan ellipses) and those SFCGS with $m_{z}$=2
(vertical violet ellipses). The HCG sample studied here consists of 91
groups where the HCG54 is not included in the graph because this
system has been considered to be a single galaxy rather than a group
\citep{Verdes-Montenegro_et_al_2002,Torres-Flores_et_al_2010}. With
respect to the SCGA, instead of choosing the catalogue B ($r'$$<$21),
we chose catalogue~A ($r'$$<$18), because this magnitude cut of
catalogue~A is comparable to the completeness limit of the GALEX/AIS
\citep{Martin_et_al_2005}.

\color{blue} The physical diameter $2r_{\rm{}phy}$ is defined here as
the double of the physical radius
$r_{\rm{}phy}$=$r_{\theta}$$\times$$d_{\theta}(z_{\rm{}UV})$ with
$d_{\theta}(z_{\rm{}UV})$ the angular-diameter distance
\citep[e.g.][]{Carroll_et_al_1992}. The angular radius $r_{\theta}$ is
computed in the same way for the four group samples considering just
the galaxies with
$\mid$$\Delta$v$_{\rm{}l-o-s}$$\mid$$\le$10$^{3}$\,km\,s$^{-1}$ in
each group. Just to give a quantitative comparison of the
2$r_{\theta}$ defined in equation \ref{eq:rtheta} with the definition
of angular diameter $\theta_{\rm{}G}$ used in \citet{Hickson_1982}, we
have checked with the 100 HCGs that the angular diameter
$\theta_{\rm{}G}$ is around 1.1 times larger than the angular diameter
2$r_{\theta}$, specifically
\mbox{$\theta_{\rm{}G}\approx[~0.9-1.3~]\times$(2$r_{\theta}$)}.

\begin{table}[!h]
\color{dreen}
\caption{\bf Comparison of significant percentiles values ($Q_{16}$,
  $Q_{50}$ and $Q_{84}$) of redshift and physical size of SFCG, HCG,
  2MCG and SCGA samples.}
\begin{center}
\begin{tabular}{c ccc }
\hline
Sample ($m_{z}$$\ge$3) & $Q_{16}$     &     $Q_{50}$  & $Q_{84}$   \\
(1) & (2) & (3) & (4) \\
\hline
 $z_{\rm{}g}$  &  &   &  \\ 
\hline
SFCG  &   0.0275 &    0.0622 &   0.0723  \\
HCG   &   0.0145 &    0.0297 &   0.0485  \\  
2MCG  &   0.0126 &    0.0201 &   0.0290  \\ 
SCGA  &   0.0615 &    0.0857 &   0.1211  \\           
\hline
 2$r_{\rm{}phy}$~(kpc) &  &    &   \\ 
\hline
SFCG  &   ~73.2  &   133.7    &  196.5 \\
HCG   &   ~57.9  &   ~89.0    &  144.8 \\
2MCG  &   106.6  &   171.6    &  275.7 \\
SCGA  &   111.4  &   164.7    &  236.4 \\
\hline
\end{tabular}
\end{center}
{\bf Columns: (1) Acronym of each compact group sample, (2)
  16-centile, (3) 50-centile and (4) 84-centile of the corresponding
  variables: $z_{\rm{}g}$ on the top and 2$r_{\rm{}phy}$ on the
  bottom.}
\label{tab:zps_comp}
\end{table}

\color{red} In Figure \ref{fig:phys_z}, the distribution in the
\mbox{2$r_{\rm{}phy}$-$z_{\rm{}UV}$} plane shows the first differences
between the compact group samples under comparison here. The SFCG
sample presents a linear trend, relatively tight, \color{dreen} in
which the nearby SFCGs have much smaller sizes (less than 100 kpc
diameters for $z$$<$0.05) compared to the distant ones (more than 200
kpc diameters for $z$$>$0.1). On the other hand, the HCG sample shows
a more dispersed relation and the 2MCG presents a more restricted
redshift range below $z$$\sim$0.05 and a size range showing the
largest values in physical size. The SCGA sample shows one major
concentration delimited by a redshift interval of 0.03$<$$z$$<$0.15
and a physical size interval of 100$<$($2r_{\rm{}phy}$/kpc)$<$250. The
differences among group samples are probably due to the specific
selection criteria applied to each sample and the original galaxy
sample from which they were extracted, as explained below.

\color{dreen} In the SFCG sample, the compactness criterion is applied
over the projected \mbox{galaxy-galaxy} angular separations of the
group members. This typically identifies groups of a similar or
smaller angular size than the linking-length value of 1.5 arcmin,
producing that, at a given angular size, distant groups have larger
physical size as one can see in Fig. \ref{fig:phys_z}. \color{blue} By
contrast, the compactness criterion of the HCG sample is applied over
the total surface brightness of group members. This search criterion
is fulfilled by galaxy groups along a sequence in which the extremes
are very compact groups of faint galaxies at one end and relatively
more extended groups of bright galaxies in the other
end. \color{dreen} Knowing that in the $2r_{\rm{}phy}$-$z_{g}$ diagram
the angular size is represented by the slope of the trend, the
dispersed relation of the HCG sample is explained the broader range of
angular sizes found for the HCG sample.

\color{red} The 2MCG sample presents a more restricted redshift range
below $z$$\sim$0.05 for being a velocity-filtered group sample mainly
retrieved from the 2MASS Redshift Survey \citep{Huchra_et_al_2012}
with a similar depth in redshift. \color{dreen} Although the 2MCG
sample is defined with the same compactness criterion, with a group
$K$-band surface brightness threshold meant to match the R-band
threshold of \citet{Hickson_1982}, the 2MCGs are somewhat less
compact. \color{red} This can be explained by the known bias of the
HCG sample towards high values of group surface brightnesses
\color{dreen} \citep{Walke&Mamon_1989,Prandoni_et_al_1994}. The SCGA
with $m_{z}$$\ge$3 is extracted from the Main Galaxy Sample (MGS) of
the Sixth Data Release \citep[DR6,][]{Adelman-McCarthy_et_al_2008} of
the Sloan Digital Sky Survey (SDSS); this explains that they are
sharing a similar redshift distribution
0.03$\lesssim$$z$$\lesssim$0.15.

\color{dreen} In Table \ref{tab:zps_comp}, one can see that the 2MCG
sample presents a distribution well concentrated around a median
redshift of $\langle$$z\rangle$=0.02, the HCG shows a bit more
extended $z$-distribution with its significant percentiles
$\sim$1.15-1.67 higher than the corresponding 2MCG redshift
percentiles. Differently, the SFCG and the SCGA samples show a clearly
more extended redshift distribution with a median redshift of
$\langle$$z\rangle$=0.062 and $\langle$$z\rangle$=0.086 greater than
the 84-centile of the two other group redshift
distributions. \color{violet} The SCGA sample is the least dispersed
of the four samples, when dispersion is measured relative to the
median.

\color{dreen} On the other hand, the physical diameter distributions
of the SFCG and HCG samples contain important fractions of groups that
are more compact than 75 kpc in physical diameter 2$r_{\rm{}phy}$. In
comparison with HCG sample, the SFCG sample contains an important part
of the population which is more extended than 130 kpc in physical
diameter 2$r_{\rm{}phy}$, increasing the median physical diameter to
$\langle$2$r_{\rm{}phy}$$\rangle$$\sim$130\,kpc in comparison with the
median physical diameter of the HCG sample,
$\langle$2$r_{\rm{}phy}$$\rangle$$\sim$90\,kpc. Differently from the
SFCG and the HCG samples, the 2MCG and the SCGA samples present a
physical diameter distribution extending towards larger values, which
is reflected in its higher percentiles compared to the two other group
samples.

\color{red} 
\subsection{Ultraviolet colour distribution and comparison with other group samples}
\label{ssec:uv_colors}

\begin{figure}[!ht]
\centering
\resizebox{1.0\hsize}{!}{\includegraphics{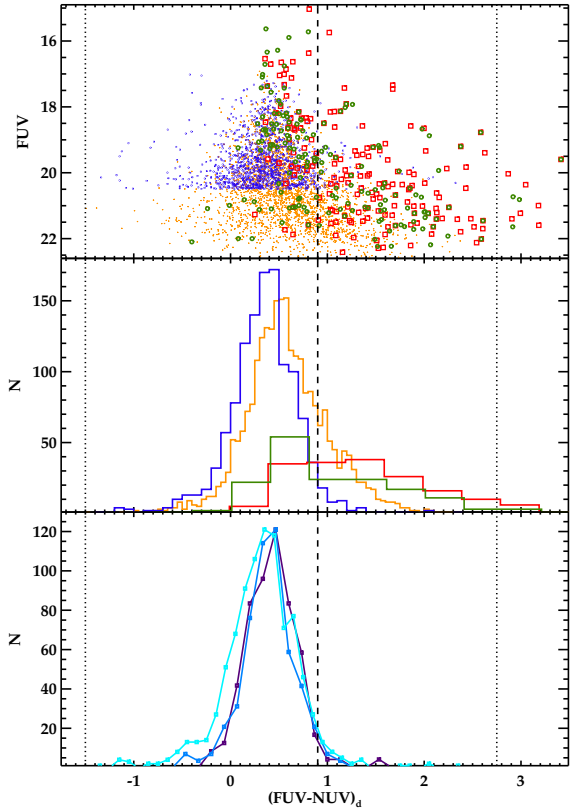}}
\caption{\bf \color{dreen} Top panel: Ultraviolet colour distributions
  of SFCG (blue), HCG (green), 2MCG (red) and SCGA (orange) samples.
  Top panel: FUV vs. (FUV$-$NUV)$_{\rm{}d}$ for the four samples.  The
  middle vertical dashed line corresponds to the FUV$-$NUV threshold
  between late-type and early-type galaxies proposed by
  \citet{GildePaz_et_al_2007}. The left and right vertical dotted
  lines delimit the colour constraint used in this work to retrieve
  ultraviolet sources from the GALEX/AIS catalogue. Middle panel:
  FUV$-$NUV color distributions of the four samples. Note that the
  histograms have different binning. Bottom panel: FUV$-$NUV color
  distributions of the subsets of galaxy members of SFCGs with
  $m_{z}$$\ge$3 (violet), galaxy members of SFCGs with $m_{z}$=2
  (blue) and non-member galaxies (cyan).  The three histograms are
  normalized to the maximum of the cyan histogram.}
\label{fig:uv_colors}
\end{figure}

Figure \ref{fig:uv_colors} displays the GALEX colour-magnitude and
color distributions for the SFCG, HCG, 2MCG and SCGA samples. In the
four samples, the panels only show those group galaxies with a
ultraviolet detection in the GALEX/AIS catalogue for both
GALEX\,bands, FUV and NUV.

\color{dreen} The panels of Figure \ref{fig:uv_colors} show three
aspects of the colour distribution of the four samples of compact
groups: the colour-magnitude diagram (top), the differential
distribution of dereddened (FUV$-$NUV)$_{\rm{}d}$ for the four group
samples (middle) and the differential distribution of three subsets of
the SFCG galaxy sample: galaxy members of SFCGs with $m_{z}$$\ge$3,
galaxy members of SFCGs with $m_{z}$=2 and the non-member galaxies
(bottom). \color{red} The three panels of Figure \ref{fig:uv_colors}
show the (FUV$-$NUV)$_{\rm{}d}$ constraints included in the search
strategy and described in Subsection \ref{ssec:uvsample} (left and
right dotted vertical lines) and also the ultraviolet colour threshold
(middle vertical dashed line) which segregates the late-types from
early-type galaxies
\citep{GildePaz_et_al_2007}. \citet{GildePaz_et_al_2007} find for a
sample of nearby galaxies that the fraction of elliptical/lenticular
galaxies with (FUV$-$NUV) colour bluer than 0.9 mag (and both FUV and
NUV magnitudes available) is 18\%, while the percentage of spiral and
irregulars redder than this value is only 12\%.

\color{red} Figure \ref{fig:uv_colors} clearly shows the limitation of
optically and near-infrared selected group samples in finding groups
dominated by star-forming galaxies, proving the interest of building a
SFCG sample because of searching over ultraviolet
catalogues. Inspecting Figure \ref{fig:uv_colors}, one can see that
virtually the entire ultraviolet colour distributions of the HCG, the
2MCG and the SCGA galaxies are within the (FUV$-$NUV)$_{\rm{}d}$
constraints included in our search strategy (depicted by the left and
right dotted vertical lines). This shows that the
(FUV$-$NUV)$_{\rm{}d}$ constraint is not the main responsible for
rejecting from our search neither optically and/or near-infrared
bright group galaxies nor early-type galaxies. In particular, the main
limitation of those group galaxies to be retrieved by the search
strategy described in this work is that they are not bright enough in
the ultraviolet spectral range. In this respect, the compact group
samples retrieved from the GALEX/AIS catalogue with the sole criterion
of being detected in both GALEX\,bands correspond \color{dreen} to
35\% of the original HCG galaxy sample of 463 galaxies, 48\% of the
original 2MCG galaxy sample of 364 galaxies and 26\% of the original
SCGA galaxy sample of 9713 galaxies, the lowest fraction of the three
samples. Once the (FUV$-$NUV)$_{\rm{}d}$ colour, photometric quality
flag and especially the FUV flux are considered in the compact group
selection, these fractions are reduced from the previous 35\% to 19\%
for the HCG galaxy sample, from the previous 48\% to 25\% for the 2MCG
galaxy sample and from the previous 26\% to 11\% for the SCGA galaxy
sample, which is again the lowest fraction of the three samples. Table
\ref{tab:uvfrac} summarizes the fractions described above for the four
samples under comparison in this work.

\begin{table}[!h]
\color{dreen}
\caption{\bf Fractions characterizing the GALEX UV magnitude and color
  distribution of the group galaxy population for each group sample.}
\begin{center}
\begin{tabular}{c ccc }
\hline
Sample & FUV\&NUV &  UV bright  & UV blue \\
   (1) &    (2)   &        (3)  & (4)     \\
\hline
SFCG   &   100    &       100   & 96      \\
HCG    &   ~35    &       ~19   & 53      \\
2MCG   &   ~48    &       ~25   & 28      \\
SCGA   &   ~26    &       ~11   & 79      \\ 
\hline
\end{tabular}
\end{center}
{\bf Columns: (1) Acronym of each compact group sample, (2) fraction
  of group galaxies detected in both GALEX bands, (3) fraction of
  group galaxies detected in both bands which fulfill the GALEX
  constraints used to select SFCG galaxies and (4) fraction of group
  galaxies detected in both bands bluer than FUV$-$NUV=0.9.}
\label{tab:uvfrac}
\end{table}

\color{dreen} On the other hand, among the galaxies that are detected
in both FUV and NUV wavebands of GALEX in the other three compact
group samples, significant fractions are bluer than the ultraviolet
colour cut proposed by \citet{GildePaz_et_al_2007}: 53\%, 28\% and
79\% for the HCG, the 2MCG and the SCGA samples, respectively (see the
bottom panel of Fig. \ref{fig:uv_colors}). Nevertheless, these very
blue galaxies are relatively less frequent in these three reference
compact group samples than they are in the SFCG sample (96\%), and the
median FUV$-$NUV is also well redder than median ultraviolet colour
$\langle$FUV$-$NUV$\rangle$=0.351 of the SFCG sample:
$\langle$FUV$-$NUV$\rangle$=0.803 for the HCG sample
$\langle$FUV$-$NUV$\rangle$=1.323 for the 2MCG sample and
$\langle$FUV$-$NUV$\rangle$=0.550 for the SCGA sample (see table
\ref{tab:uvd_comp}).

\color{dreen} In the bottom panel of Fig. \ref{fig:uv_colors}, it is
also displayed the UV colour distributions of galaxy members of SFCGs
with $m_{z}$$\ge$3 (violet lines), $m_{z}$=2 (blue lines) and also for
the non-member galaxies (cyan lines) with the aim of checking the
presence of a population of interlopers with a different ultraviolet
colour distribution.

\begin{table}[!h]
\color{dreen}
\caption{\bf Comparison of significant percentiles values ($Q_{16}$,
  $Q_{50}$ and $Q_{84}$) of UV colors of SFCG, three subsamples of the
  former sample, HCG, 2MCG and SCGA samples.}
\begin{center}
\begin{tabular}{c cccc }
\hline
 (FUV$-$NUV)$_{\rm{}d}$ &    &  &  &  \\
\hline
Sample       & $Q_{16}$  &  $Q_{50}$  & $Q_{84}$ &   N     \\
(1)          & (2)      & (3)       & (4)     &  (5)    \\
\hline 
SFCG          &   0.048   &  0.351  &   0.652  &  1193  \\
$m_{z}$$\ge$3  &   0.180   &  0.421  &   0.669  &  ~128  \\
$m_{z}$=2      &   0.146   &  0.369  &   0.635  &  ~148  \\
non-membs     &   0.014   &  0.332  &   0.652  &  ~904  \\
HCG           &   0.406   &  0.803  &   1.853  &  ~162  \\
2MCG          &   0.649   &  1.323  &   2.123  &  ~174  \\
SCGA          &   0.208   &  0.550  &   0.990  &  2517  \\            
\hline
\end{tabular}
\end{center}
{\bf Columns: (1) Acronym of each compact group sample, (2)
  16-centile, (3) 50-centile, (4) 84-centile of the
  (FUV$-$NUV)$_{\rm{}d}$ distribution and (5) the total number of
  galaxies detected in both GALEX bands.}
\label{tab:uvd_comp}
\end{table}

A Kolmogorov-Smirnov test applied to the $m_{z}$=2 and $m_{z}$$\ge$3
UV color distributions produces a probability of 32\% that the two
colour distributions come from the same parent population. The same
test applied between the UV color distributions of the non-member
population and the $m_{z}$=2 population gives a 1.3\% probability
whereas this test gives a 0.01\% probability between the non-member
population and the $m_{z}$$\ge$3 population. Therefore, the non-member
population is statistically distinguishable from the $m_{z}$=2 and
$m_{z}$$\ge$3 populations, which are very similar between them. This
fact can be also observed in the bottom panel of
Fig. \ref{fig:uv_colors} \color{violet} and the second and third rows
of Table \ref{tab:uvd_comp} \color{dreen} in the comparison of the UV
color distributions of these three SFCG subsamples where there is an
extra population of galaxies with FUV$-$NUV$\lesssim$0 in the
subsample of non members.

\color{dreen} To analyse the origin of this extra population having
FUV$-$NUV$\lesssim$0 in the non-member subsample, we have computed the
k-correction using the calculator provided by
\citet{Chilingarian&Zolotukhin_2011}\footnote{http://kcor.sai.msu.ru}. We
computed the difference between the observed GALEX colour
(FUV$-$NUV)$_{\rm{}obs}$ and the rest-frame GALEX colour
(FUV$-$NUV)$_{\rm{}r-f}$ as a function of redshift for blue galaxies
i.e: \\

(FUV$-$NUV)$_{\rm{}obs}$ $-$ (FUV$-$NUV)$_{\rm{}r-f}$  = \hfill 
\begin{equation}
\hfill = - [~k_{\rm{}FUV}(z;~{\rm color}_{1}) - k_{\rm{}NUV}(z;~{\rm color}_{2})~].   
\label{eq:uvkcorr}
\end{equation}

The k-correction $k_{\lambda}$($z$; color$_{i}$) is defined as
$k_{\lambda}$($z$;~color$_{i}$)=$m_{\rm{}r-f}$$-$$m_{\rm{}obs}$($z$). The
k-correction is applied to the FUV\,band taking a GALEX color for blue
galaxies of color$_{1}$$\equiv$FUV$-$NUV$\sim$0.35 (see
Fig. \ref{fig:uv_colors}) and the k-correction is applied to the
NUV\,band assuming a SDSS-GALEX color for blue galaxies of
color$_{2}$$\equiv$NUV$-$$r'$$\sim$2
\citep[c.f.][]{Hernandez-Fernandez_et_al_2012b}. We find that the
(FUV$-$NUV)$_{\rm{}obs}$ of blue galaxies as a function of redshift
only goes bluer than the (FUV$-$NUV)$_{\rm{}r-f}$ for redshifts above
$z$$\sim$0.1. This suggests that the bluest FUV$-$NUV colours are
indeed caused by galaxy interlopers or also galaxy groups at
$z$$>$0.1.

Comparing the 16-centiles of the UV color distributions of the SFCG
subsets, $m_{z}$$\ge$3, $m_{z}$=2 and non-members (see Table
\ref{tab:uvd_comp}), we should note that the 16-centile of the
non-member galaxy sample $Q_{\rm{}16}$=0.014 is clearly lower than the
other two subsets, $Q_{\rm{}16}$=0.146 for $m_{z}$$\ge$3 and
$Q_{\rm{}16}$=0.180 for $m_{z}$=2. This suggests that interlopers play
a role in producing even bluer median colors, but that this effect
$Q_{\rm{}50}$(SFCG)$-$$Q_{\rm{}50}$($m_{z}$$\ge$3)=$-$0.07 is much
smaller than the difference in median colors with the 2MCG, HCG and
SCGA samples, $Q_{\rm{}50}$(SCGA)$-$$Q_{\rm{}50}$($m_{z}$$\ge$3)=0.129
in the worst case.

\color{blue}
\subsection{Results from an ultraviolet search over the galaxy catalogue of HCG sample}
\label{ssec:uvhick}

With the aim of comparing the approach of identifying galaxy groups
using catalogues of UV emitting galaxies with the approach of
searching groups in catalogues of optically and near-infrared selected
galaxies, we apply the same methodology described in this work over
the sample of group galaxies from the HCG catalogue
\citep{Hickson_et_al_1992} consisting of 463 galaxies. We take the HCG
sample as the prototype of the optically or near-infrared selected
group samples, assuming that similar results should be expected for
others \citep[e.g.][]{
  Prandoni_et_al_1994,Barton_et_al_1996,Diaz-Gimenez_et_al_2012}. The
test provides the following results:

\begin{itemize}

\item From a total of 463 HCG galaxies, only 87 galaxy members fulfill
  the constraints on FUV brightness, UV colour and GALEX photometric
  quality. This means that on average, the HCG sample contains less
  than one UV-bright galaxy per group.

\item Specifically, only three HCGs have four or more HCG galaxy
  members fulfilling the GALEX UV constraints: HCG23, HCG89 and HCG100
  with four UV-bright galaxy members in each of them. This low ratio
  (3\%) points out the main limitation of the HCGs to be included in a
  sample of SFCGs; they are mainly populated by galaxies without a
  bright UV counterpart.

\item Furthermore, there are no HCGs with four or more UV-bright
  members and also fulfilling our compactness criterion: a maximum
  angular distance of 1.5 arcmin from each member to some of the rest
  of HCG galaxy members.

\item The output of this test just gives HCG100 as a compact group of
  UV-bright galaxies but only with three UV-bright members close
  enough. Then, HCG100 should not be in principle included in the SFCG
  sample. However, HCG100 (identified here as SFCG2) does belong to
  the sample of SFCGs because there is a fourth UV-bright galaxy not
  in the HCG catalogue, \color{dreen} which lies
  \mbox{$\approx$\,1\,arcmin} to the Southeast of HCG100b
  \citep[denoted HCG100x
    by][]{Plana_et_al_2003}\footnote{\color{dreen} The celestial
    coordinates of this object are R.A.=0:01:29.52,
    Dec.=13:05:54.9}. \color{red} This galaxy HCG100x is therefore a
  member of the selected SFCG, while the HCG member HCG100c does not
  fulfill the compactness criterion applied in this work, a maximum
  angular separation of 1.5 arcmin (see the corresponding image in the
  stamp collection).

\item \color{red} We note that there is one of the HCG sample that
  deserves maintaining in this context as an extreme case, the system
  HCG54. This system has been considered to be a single galaxy rather
  than a group
  \citep{Verdes-Montenegro_et_al_2002,Torres-Flores_et_al_2010}. HCG54
  is identified as one single UV source by the GALEX photometric code.

\end{itemize}

\color{dreen} 
\subsection{Kinematical analysis and comparison with other group samples}
\label{ssec:kyn}

\begin{figure}[!hb]
\centering
\resizebox{1.00\hsize}{!}{\includegraphics{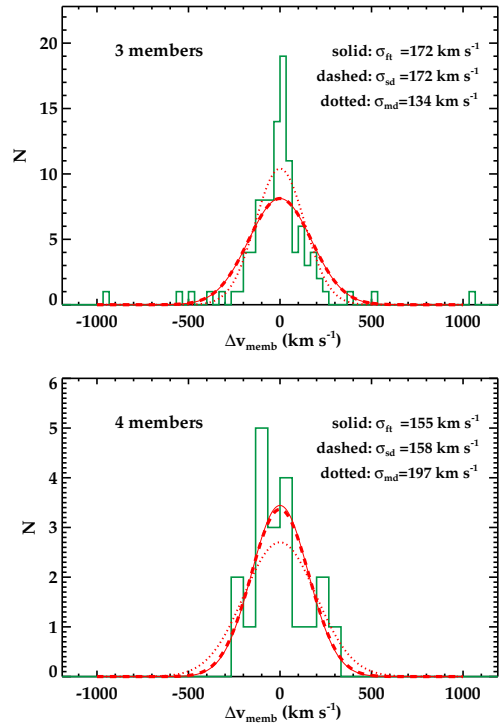}}
\caption{\bf \color{dreen} Distributions of line-of-sight velocities
  (relative to the group mean) of the subset of SFCGs with $m_{z}$=3
  (top panel) and $m_{z}$=4 (bottom panel). Solid curves are fits to
  the stacked distribution of velocities relative to the group mean
  velocity. Dashed and dotted curves show the normal distributions
  with standard deviations equal to that of the $\Delta$v distribution
  and to the median of the standard deviations, respectively. The
  l-o-s velocity dispersion values derived from the fitting
  $\sigma_{\rm{}ft}$, from the standard deviation of the
  $\Delta$v$_{\rm{}memb}$ distribution $\sigma_{\rm{}sd}$ and the
  median velocity dispersion of the corresponding subsample
  $\sigma_{\rm{}md}$ are shown at the upper left region of each
  panel.}
\label{fig:vdisp}
\end{figure}

Figure \ref{fig:vdisp} shows results of a kinematical analysis of the
subset of SFCGs with more than three members (top panel, three members
$m_{z}$=3 and bottom panel, four members $m_{z}$=4). This comprises 36
groups with $m_{z}$=3 and 5 groups with $m_{z}$=4. Instead of showing
the distribution of l-o-s velocities with respect to the median
redshift i.e. $\Delta$v$_{\rm{}l-o-s}$, we show the velocity
distribution with respect to the average redshift
$\langle$$z_{\rm{}memb}$$\rangle$ of the $m_{z}$ galaxy members i.e.:

\begin{equation}
\Delta{\rm v}_{\rm{}memb}={\rm c}~\left(\frac{z_{\rm{}memb}-\langle{}z_{\rm{}memb}\rangle}{1+\langle{}z_{\rm{}memb}\rangle}\right).
\label{eq:dvmemb}
\end{equation}

This avoids the artifact of a central peak with
$\Delta$v$_{\rm{}l-o-s}$=0. \color{orange} The solid lines depict the
normal distributions derived from the maximum likelihood fitting,
i.e. by minimizing

\begin{equation}
-\ln\mathcal{L}=-\sum\limits_{i}\ln p_{i}, 
\end{equation}

where $\mathcal{L}=\prod\limits_{i}p_{i}$ is the likelihood,

\begin{equation}
p_{i}=\frac{1}{\sqrt{2\pi\sigma^{2}}} \exp \left[-\frac{(\Delta\rm{v}_{\rm{}memb} - \mu)^{2}}{2\sigma^{2}}\right], 
\end{equation}

and where $\Delta\rm{v}_{\rm{}memb}$ is the velocity offset of galaxy
{\it i} and $\mu$ is the mean velocity offset from the group mean
velocities. Naturally, it makes no sense to consider non-zero mean
velocity offsets from the group mean velocities, so we simplify to
$\mu$=0. \color{brown} The dashed lines show the normal distributions
fixed to the standard deviation directly derived from the
$\Delta$v$_{\rm{}memb}$ distribution. The dotted lines show the normal
distributions fixed to the median velocity dispersion of the
corresponding subsample.

\color{dreen}

\color{orange} The derived values of l-o-s velocity dispersion for
those $m_{z}$=3 SFCGs are in the range of
$\sigma_{\rm{}l-o-s}$$\approx$135-170\,km\,s$^{-1}$ which contains the
median velocity dispersion of the spiral-rich HCGs
($\sigma_{\rm{}l-o-s}$$\approx$140\,km\,s$^{-1}$) but it is clearly
lower than the median $\sigma_{\rm{}l-o-s}$ of the spiral-poor HCGs
\citep[$\sigma_{\rm{}l-o-s}$$\gtrsim$250\,km\,s$^{-1}$,][]{Hickson_1997}. In
the case of SFCGs with $m_{z}$=4, the derived values of l-o-s velocity
dispersion are in the range of
$\sigma_{\rm{}l-o-s}$=155-200\,km\,s$^{-1}$ which is closer to the
median $\sigma_{\rm{}l-o-s}$ of the spiral-rich HCGs
\citep[][fig. 3]{Hickson_1997}. \color{red} The fact that the typical
SFCG velocity dispersion is closer to that of the spiral-rich HCGs
\citep{Hickson_1997} is in agreement with the high fraction of
late-type galaxies expected from the FUV$-$NUV colours shown by the
SFCG galaxy sample i.e. FUV$-$NUV$\lesssim$0.9.

\begin{figure}[!ht]
\centering
\resizebox{1.00\hsize}{!}{\includegraphics{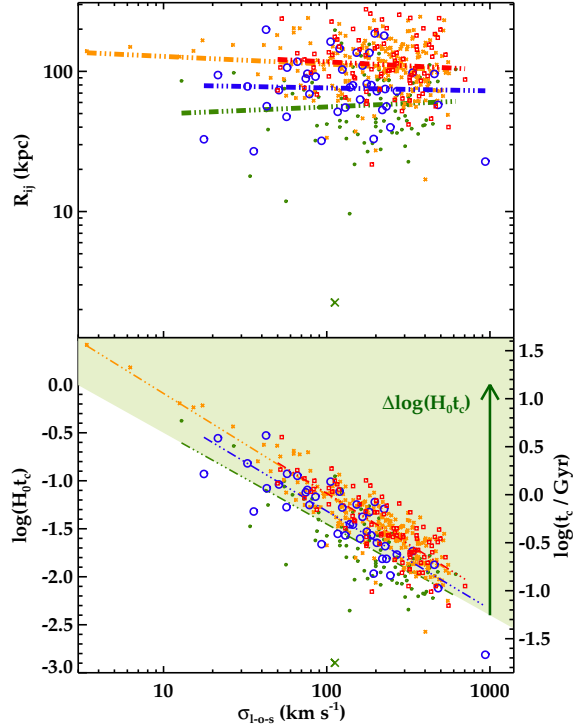}}
\caption{\bf \color{dreen} Blue empty circles, green points, red
  squares and orange crosses correspond to the SFCG, HCG, 2MCG and
  SCGA samples, respectively. HCG54 is depicted by a green cross at
  the bottom of each panel. Linear fits for each sample is shown with
  the same color code. Top panel: Median projected physical separation
  vs. l-o-s velocity dispersion. Bottom panel: Crossing time vs. l-o-s
  velocity dispersion. The crossing time $t_{\rm{}c}$ is shown in
  dimensionless (left axis) and physical (right axis) units. Green
  shaded area delimits the diagram space where
  $\Delta$$\log$($H_{\rm{}0}$$t_{\rm{}c}$) (see Eq. \ref{eq:DlHt}) is
  considered positive in the increasing direction indicated by the
  vertical green arrow.}
\label{fig:vdisp1}
\end{figure}

\color{dreen} The top panel of Figure \ref{fig:vdisp1} shows the
distribution of the median projected physical separations $R_{ij}$
with respect to the velocity dispersion $\sigma_{\rm\,l-o-s}$ for the
sample of SFCGs (blue empty circles), the HCG sample (green points),
the 2MCG sample (red squares) and the SCGA sample (orange
crosses). The median projected physical separation $R_{ij}$ is
computed as the median physical length of the two-dimensional
galaxy-galaxy separation vectors of the galaxy members for the SFCGs,
HCGs and SCGAs with $m_{z}$$\ge$3. The $R_{ij}$ values of the 2MCG
sample are translated from the original work
\citep{Diaz-Gimenez_et_al_2012} to the cosmological model used in this
work. \color{brown} The top panel also includes the four linear fits
in the log-log plane of these variables i.e.:

$\log(R_{ij}/{\rm{}kpc})_{\rm{}fit}=A\log(\sigma_{\rm{}l-o-s}/{\rm{}km\,s}^{-1})+B$. 

\color{dreen} The parameter values derived from these linear fits and
their associated uncertainties are shown in Table \ref{tab:lrlsfit}.

\begin{table}[!h]
\color{dreen}
\caption{\bf Derived parameters from linear fitting in the
  $\log$($R_{ij}$) vs. $\log$($\sigma_{\rm{}l-o-s}$) plane.}
\begin{center}
\begin{tabular}{c cc }
\hline
Sample & $A\pm\delta$$A$ & $B\pm\delta$$B$ \\
(1) & (2) & (3) \\
\hline
SFCG  & $-$0.021  $\pm$   0.100   &    1.92~  $\pm$ 0.21~ \\   
\color{orange} HCG   &  \color{orange}~~0.050  $\pm$   0.082   &   \color{orange} 1.65~  $\pm$ 0.19~ \\ 
2MCG  & $-$0.058  $\pm$   0.086   &    2.18~  $\pm$ 0.20~ \\   
SCGA  & $-$0.055  $\pm$   0.036   &    2.162  $\pm$ 0.082 \\   
\hline
\end{tabular}
\end{center}
{\bf Columns: (1) Acronym of each compact group sample, (2)
  logarithmic slope and (3) ordinate at the origin of the linear fit
  $\log(R_{ij}/{\rm{}kpc})_{\rm{}fit}=A\log(\sigma_{\rm{}l-o-s}/{\rm{}km\,s}^{-1})+B$.}
\label{tab:lrlsfit}
\end{table}

The top panel of Figure \ref{fig:vdisp1} shows clearly that the
different group samples present comparatively distinct size averages
for any given velocity dispersion. Despite the four group samples are
distributed in a similar range of $R_{ij}$,
20$\lesssim$($R_{ij}$/kpc)$\lesssim$200, their medians (see Table
\ref{tab:st_comp}) show distinct values: $R_{ij}$$\sim$80 kpc for the
SFCGs, \textcolor{orange}{$R_{ij}$$\sim$60 kpc} for the HCG and
$R_{ij}$$\sim$110 kpc for the 2MCG and SCGA. Further, the linear fits
$\log$($R_{ij}$) vs. $\log$($\sigma_{\rm{}l-o-s}$) presents a
negligible slope within uncertainties (see Table
\ref{tab:lrlsfit}). This fact can be interpreted as an effect of the
selection of groups in projection which is not biased toward any of
the situations characterizing groups selected in projection:
physically rounded groups, groups physically elongated along the
line-of-sight direction or also chance galaxy alignments along the
line-of-sight direction. The predominance of each one of these
situations has been analysed by \citet{Diaz-Gimenez&Mamon_2010}.

We should add that, according to the median values of velocity
dispersion and median projected physical separation of the four
compact group samples given in Table \ref{tab:st_comp} and assuming
the virial condition i.e $M_{\rm{}vir}$$\propto$$\sigma^{2}$$R_{ij}$
where $M_{\rm{}vir}$ corresponds to the virial theorem mass of the
system, the virial theorem masses of SFCGs are on average
\textcolor{orange}{1.6}, 4.2 and 3.8 times lower than the HCG, 2MCG
and SCGA samples, respectively.

\begin{table}[!h]
\color{dreen}
\caption{\bf Comparison of significant percentiles values ($Q_{16}$,
  $Q_{50}$ and $Q_{84}$) of velocity dispersion, crossing time and
  median projected physical separation of SFCG, HCG, 2MCG and SCGA samples.}
\begin{center}
\begin{tabular}{c ccc }
\hline
Sample ($m_{z}$$\ge$3) & $Q_{16}$ & $Q_{50}$ & $Q_{84}$ \\
(1) & (2) & (3) & (4) \\
\hline
 $\sigma_{\rm{}l-o-s}$~(km\,s$^{-1}$) & & & \\
\hline
SFCG  &   ~51.0   &   138.7   &  228.0  \\  
HCG   &   ~89.1   &   204.2   &  323.6  \\  
2MCG  &   108.0   &   241.0   &  391.0  \\  
SCGA  &   100.3   &   227.1   &  340.0  \\          
\hline
 log($H_{0}$$t_{\rm{}c}$) & & & \\
\hline 
SFCG  &   -1.81   &   -1.33 &   -1.04 \\   
HCG   &   -2.03   &   -1.78 &   -1.33 \\   
2MCG  &   -1.82   &   -1.55 &   -1.09 \\   
SCGA  &   -1.82   &   -1.48 &   -1.13 \\       
\hline
$R_{ij}$~(kpc) &  &  &  \\
\hline
SFCG  &  47.4   &   ~78.1  &    117.2  \\   
\color{orange}HCG   &  \color{orange}36.3   &   \color{orange}~58.7  &    \color{orange}~94.5  \\   
2MCG  &  68.4   &   108.3  &    180.3  \\  
SCGA  &  72.3   &   111.5  &    157.9  \\  
\hline
\end{tabular}
\end{center}
{\bf Columns: (1) Acronym of each compact group sample, (2)
  16-centile, (3) 50-centile and (4) 84-centile of the corresponding
  variable: $\sigma_{\rm{}l-o-s}$ on the top,
  $\log$($H_{\rm{}0}$$t_{\rm{}c}$) in the middle and $R_{ij}$ on the
  bottom.}
\label{tab:st_comp}
\end{table}

\color{dreen} Figure \ref{fig:vdisp1} also shows in its bottom panel
the relation of the crossing-time $t_{\rm{}c}$ with respect to the
l-o-s velocity dispersion $\sigma_{\rm{}l-o-s}$ (bottom panel) for the
four samples. The crossing-time is defined as:

\begin{equation}
t_{\rm{}c}=\frac{\pi}{2}\frac{R_{ij}}{\sqrt{3}~\sigma_{\rm{}l-o-s}}.
\label{eq:tc}
\end{equation}

\color{blue} The crossing time values from the 2MCG sample are
translated to the cosmological model used in this work whereas the
crossing time values of the SFCG, HCG and SCGA samples are computed
with the very same cosmological model and with the crossing time
formula given in Equation \ref{eq:tc}. \color{brown} The bottom panel
of Figure \ref{fig:vdisp1} also displays the linear fits in the
log-log plane i.e.

$\log(H_{\rm{}0}t_{\rm{}c})_{\rm{}fit}=M\log(\sigma_{\rm{}l-o-s}/{\rm{}km\,s}^{-1})+N$,

to the four samples. \color{blue} The parameter values derived from
these linear fits and their associated uncertainties are shown in
Table \ref{tab:ltlsfit}. The linear fit to the HCG sample does not
include the HCG54 with a crossing time value
$\log$($H_{\rm{}0}$$t_{\rm{}c}$)$\approx$$-$2.9 because it is showing
a clear outlier behaviour and it is considered a galaxy remnant and
not a galaxy group
\citep{Verdes-Montenegro_et_al_2002,Torres-Flores_et_al_2010}.

\begin{table}[!h]
\color{dreen}
\caption{\bf Derived parameters from linear fitting in the
  log($H_{\rm{}0}$$t_{\rm{}c}$) vs. log($\sigma_{\rm{}l-o-s}$) plane.}
\begin{center}
\begin{tabular}{c cc }
\hline
Sample & $M\pm\delta$$M$ & $N\pm\delta$$N$ \\
(1) & (2) & (3) \\
\hline
SFCG  & $-$1.021 $\pm$ 0.100 & 0.73~ $\pm$ 0.21~  \\  
 HCG  & $-$0.950 $\pm$ 0.083 & 0.45~ $\pm$ 0.19~  \\  
2MCG  & $-$1.056 $\pm$ 0.087 & 0.98~ $\pm$ 0.20~  \\  
SCGA  & $-$1.055 $\pm$ 0.036 & 0.964 $\pm$ 0.082  \\                     
\hline
\end{tabular}
\end{center}
{\bf Columns: (1) Acronym of each compact group sample, (2)
  logarithmic slope and (3) ordinate at the origin of the linear fit
  $\log(H_{\rm{}0}t_{\rm{}c})_{\rm{}fit}=M\log(\sigma_{\rm{}l-o-s}/{\rm{}km\,s}^{-1})+N$.}
\label{tab:ltlsfit}
\end{table}

\color{blue} The linear fits in Fig. \ref{fig:vdisp1} present similar
slopes for the four samples and they are close to minus one within
uncertainties. If the fitted slopes are not affected by incompleteness
in the plane
\mbox{log($H_{\rm{}0}$$t_{\rm{}c}$)-log($\sigma_{\rm{}l-o-s}$)}, this
coincidence of $M$$\approx$$-1$ means that the projected physical sizes
of compact groups in these samples would not present a significant
trend with velocity dispersion and it can be seen also in the top
panel of Figure \ref{fig:vdisp1}. This is contrary to one would expect
for a galaxy system assuming a principle of homology.

\color{dreen} On the other hand, the 2MCG and SCGA samples, which
share a very similar trend in the log($H_{\rm{}0}$$t_{\rm{}c}$)
vs. log($\sigma_{\rm{}l-o-s}$) plane, present systematically longer
crossing times than the SFCG sample and the SFCG sample shows
systematically longer crossing than the HCG sample, for a given
velocity dispersion. This places the four group samples in an
evolutionary sequence from a dynamical point of view, in which the HCG
sample is the more dynamically evolved group sample and the SFCG
sample is in an intermediate stage between the HCGs at one end, and
the 2MCG and SCGA samples in the other end, being these ones less
dynamically evolved.

\color{brown} Specifically, for given group velocity dispersion, the
typical crossing time of the SFCG sample with respect to the HCG
linear fit (i.e.~$\log(H_{\rm{}0}t_{\rm{}c})^{\rm{}HCG}_{\rm{}fit}$)
is on average 36\% longer
i.e.~$\Delta$$\log$($H_{\rm{}0}$$t_{\rm{}c}$)=0.13, where we define
$\Delta$$\log(H_{\rm{}0}t_{\rm{}c})$ as:

\begin{equation}
\Delta\log(H_{\rm{}0}t_{\rm{}c})=\log(H_{\rm{}0}t_{\rm{}c})-\log(H_{\rm{}0}t_{\rm{}c})^{\rm{}HCG}_{\rm{}fit}.
\label{eq:DlHt}
\end{equation}

This crossing time increase reaches up to a 100\%
(i.e.~$\Delta$$\log$($H_{\rm{}0}$$t_{\rm{}c}$)=0.30) for the 2MCG and
the SCGA samples with respect to the HCG sample.

\begin{figure}[!ht]
\centering
\resizebox{1.00\hsize}{!}{\includegraphics{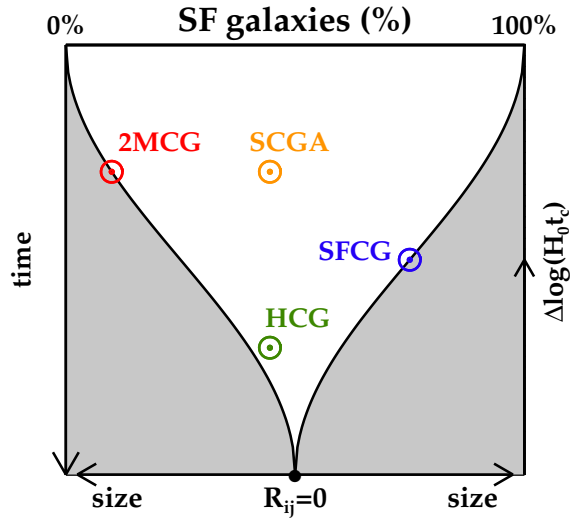}}
\caption{\bf \color{red} Scheme which summarizes and compares the
  typical characteristic evolutionary stage, fraction of star-forming
  galaxies and physical size of the three group samples compared
  here. \color{violet} The vertical axis depicts an evolutionary
  sequence (left axis) in which the proxy is considered the surplus in
  crossing time with respect to the HCG linear fit
  $\Delta$$\log$($H_{\rm{}0}$$t_{\rm{}c}$) (right axis), see
  Eq. \ref{eq:DlHt}. \color{red} The horizontal axis attempts to
  account for the star-forming fraction of galaxy members at the top
  of the graph and, the evolution of the physical size of the galaxy
  groups along cosmic time, at the bottom.}
\label{fig:sfcg_squema}
\end{figure}

\color{blue} For the SFCGs, the larger crossing times with respect to
the HCG sample can be interpreted as an indication of less dynamically
evolved groups and therefore, that their galaxies have been less
affected by galaxy-galaxy interactions and environmental processes
linked to the intragroup plasma and the dark matter group halo. This
would allow the presence of a significant pristine gas reservoir in
group galaxies and consequently such abundance of star-forming
galaxies in the SFCGs. This does not seem to be the situation for the
2MCG and the SCGA samples as these groups have been extracted from
near-infrared and optical selected galaxy catalogues respectively and,
as shown in Subsections \ref{ssec:uv_colors} and \ref{ssec:uvhick},
the 2MCG and the SCGA samples should be poorly populated by
star-forming galaxies. We outline a comprehensive explanation for both
situations. Galaxy groups can be assembled from a mixture of
previously passive and/or previously star-forming galaxies. Group
searches in the near-infrared and optical catalogues are finding those
groups mainly assembled from passive galaxies whereas the searches in
ultraviolet catalogues are finding those groups mainly assembled from
previously star-forming galaxies which still retain a significant
amount of fresh gas. \color{dreen} The evolutionary scenario
previously described is graphically summarized in Figure
\ref{fig:sfcg_squema}.

\color{black}
\section{Cross-matching with compiled catalogues of galaxies, groups and galaxy clusters}
\label{sec:x-matching}

\subsection{Search for analogues of BIG}
\label{ssec:big}

\begin{figure}[!ht]
\centering
\resizebox{1.00\hsize}{!}{\includegraphics{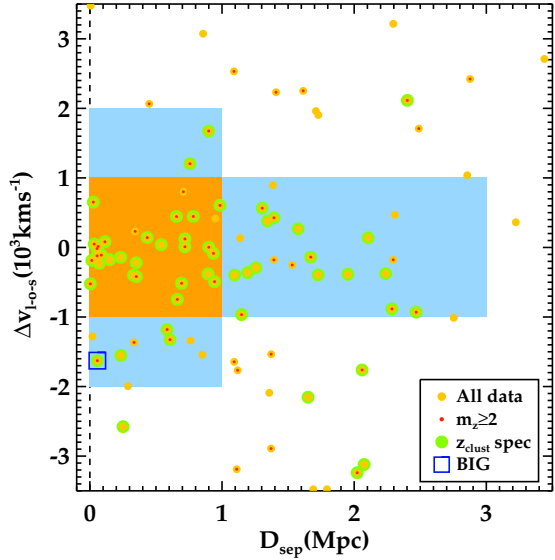}}
\caption{\bf Stack of projected phase space diagram of SFCGs with
  respect to their candidate parent cluster, $\Delta$v vs. $D_{\rm
    sep}$. The red points represent those groups with
  $m_{\rm{}z}$$\ge$2. Green large circles mark those groups where the
  redshift of the candidate parent cluster is
  spectroscopically-derived indicating an accurate cluster
  redshift. The BIG is labeled by a blue square. Those groups
  tentatively identified as infalling groups are those ones located
  over the blue shaded area while the groups considered as groups
  located in the central region of clusters are those ones located
  over the orange shaded area.}
\label{fig:cpps_big}
\end{figure}

With the aim of finding analogues to the Blue Infalling Group, the
better known example of a SFCG infalling to a cluster (ABELL\,1367),
we cross-matched the sample of SFCGs with the largest compilation of
catalogues of galaxy clusters, NED (NASA/IPAC Extragalactic
Database). We searched for objects compiled by NED as galaxy clusters
(``GClust") up to a searching radius of 50 arcmin. We set the upper
redshift to $z$=0.3, noting that the highest values of SFCG redshifts
are around $z$$\sim$0.15. Figure \ref{fig:cpps_big} shows the
distribution of the resulting SFCG-cluster matches in projected
physical separation $D_{\rm{}sep}$ in Mpc, and line-of-sight velocity
difference $\Delta$v, with respect to the candidate parent
cluster. Note that both cluster virial radii and/or velocity
dispersions are not known or computed for every cluster compiled by
NED.

\color{blue} The precise definition of the infalling regions in the
cluster projected phase has been recently studied
\citep[e.g.][]{Mahajan_et_al_2011b,Haines_et_al_2012,Oman_et_al_2013}
and always suffers of overlapping problems because a six-dimensional
phase space is being collapsed into a two dimensional projected phase
space. \color{red} Because of these limitations, we just tentatively
identify as infalling groups those that are sufficiently close to the
parent cluster i.e.:

\color{red}
$D_{\rm{}sep}$$<$1\,Mpc {\bf\,AND}  1000$<$$\left(\frac{\mid\Delta \rm v\mid}{\rm km\,s^{-1}}\right)$$<$2000

{\bf\,OR}

1$<$$\left(\frac{D_{\rm{}sep}}{\rm Mpc}\right)$$<$3 {\bf\,AND} $\mid$$\Delta$v$\mid$$<$1000\,km\,s$^{-1}$ \\

\color{blue} but excluding the cluster central regions:

$D_{\rm{}sep}$$<$1\,Mpc {\bf\,AND} $\mid$$\Delta$v$\mid$$<$1000\,km\,s$^{-1}$. \\

\color{dreen} Within these boundaries, we find 26 SFCGs embedded in the
infall regions of 32 candidate parent clusters.  Note that a few of
the SFCGs are embedded in the infall regions of more than one
cluster. We also find 31 SFCGs in the central regions of
clusters. In this cluster central region, most of the SFCGs should be
identified with the cluster itself.

\color{black}

\subsection{Cross-matching with catalogues of galaxy groups and galaxies}
\label{ssec:xmatchz}

\subsubsection{Cross-matching with catalogues of known galaxy groups}

In order to know whether there is a previously identified galaxy group
at the sky position of each SFCG, we tentatively crossmatch the
sample with those objects classified by NED as galaxy clusters
(``GClust") or galaxy groups (``GGroup") inside a searching radius of
6 arcmin. Knowing that the linking length in the FoF algorithm is set
to a maximum of 1.5 arcmin, we can assume that a searching radius of 6
arcmin is enough to find a galaxy group counterpart for each SFCG, if
it exists. The redshifts (when available) of NED groups are considered
as additional estimates for the matching SFCGs.

\begin{figure}[!ht]
\centering
\resizebox{0.975\hsize}{!}{\includegraphics{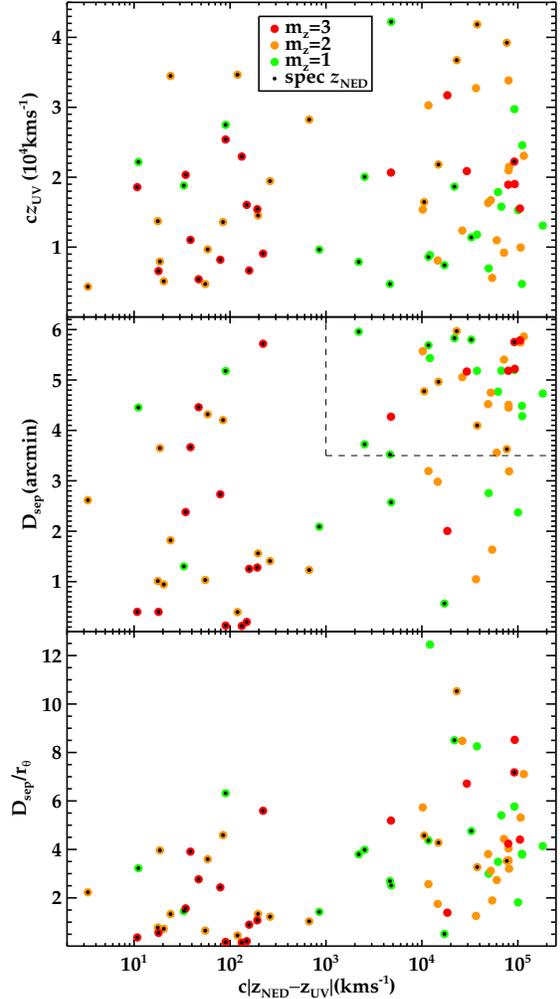}}
\caption{\bf Diagnostics of matches between SFCGs and NED
  group/clusters of known redshift in terms of absolute l-o-s velocity
  difference. {\it Top}: l-o-s velocity of SFCGs. Colour code
  identifies the numbers of known UV-galaxy redshifts to define the
  SFCG redshift. The black points mark those CSGFs where the redshift
  of the NED group is spectroscopically-derived. {\it Middle}: Angular
  separation. The two dashed lines at
  $\mid$$\Delta$v$\mid$=1000\,km\,s$^{-1}$ and
  $D_{\rm{}sep}$=3.5\,arcmin indicate those groups with large velocity
  differences and separation distances, respectively (see text for
  details). {\it Bottom}: Angular separation in units of SFCG size
  $r_{\rm{}\theta}$.}
\label{fig:group-xm1}
\end{figure}

As can be seen in the three panels of Figure \ref{fig:group-xm1}, two
sets of crossmatched group pairs are clearly distinguishable, those
with velocity differences lower than
$\mid$$\Delta$v$\mid$$\sim$1000~km\,s$^{-1}$ and those with larger
velocity differences. Inspecting the subset with large velocity
differences, we note that there are six examples of galaxy groups with
more robust group redshifts $z_{\rm{}UV}$ i.e. $m_{z}$$\ge$2. These
ones correspond to the red and orange data circles with a central
black point in Fig. \ref{fig:group-xm1}. A first characteristic of
this subset (all in the upper right portion of the middle panel of
Figure \ref{fig:group-xm1}) is that it presents groups with large
separations ($D_{\rm{}sep}$$>$3.5~arcmin), suggesting that there is no
physical connection between the SFCG and the previously catalogued
group. This is confirmed by the $D_{\rm{}sep}$/$r_{\theta}$ values
(see Table \ref{tab:suspg} and bottom panel in figure
\ref{fig:group-xm1}) which show that \textcolor{blue}{the centers of
  the NED groups lie more than three SFCG group radii away from these
  SFCG groups and in consequence, clearly outside from the physical
  extension of the SFCG.} These groups and their corresponding NED
group counterparts are shown in Table \ref{tab:suspg} and they are
described in detail in the following:

\onecolumn
\begin{table}[!hb]
\caption{\bf SFCGs with large velocity differences and large
  separations with respect to their known group counterparts.}
\begin{center}
\resizebox{1.00\hsize}{!}{\bf
\begin{tabular}{lll l lll ll}
\hline
$n_{\rm{}ID}$&   Object Name    &$z_{\rm{}NED}$ &$\Delta$v  & RA       &  Dec        &$D_{\rm{}sep}$&$r_{\theta}$   \\
         &                   &           &km\,s$^{-1}$ &  deg       &  deg        & arcmin    & arcmin       \\
  (1)    & (2)               &  (3)      & (4)       & (5)         &  (6)        & (7)       & (8)          \\
\hline
   92    &  SDSSCGB 46521    & 0.3810    &  ~91986   &  161.708    &  43.0632    &  5.7      &  0.801       \\
  106    &  WBL 340          & 0.0196    &  -10560   &  173.910    &  54.8901    &  4.8      &  1.045       \\
  117    &  ABELL 1514       & 0.1995    &  ~23074   &  184.484    &  20.6542    &  6.0      &  0.567       \\
  134    &  $[$RPG97$]$ 236  & 0.0232    &  -14864   &  198.638    &  17.1708    &  5.0      &  1.160       \\
  146    &  SDSSCGB 49472    & 0.3870    &  ~76793   &  206.894    &  32.5728    &  3.6      &  1.029       \\
  256    &  UGCl 314         & 0.0139    &  -37688   &  211.808    &  12.6627    &  4.1      &  1.253       \\ 
\hline
\end{tabular}}
\end{center}
{\bf (1) ID number of the SFCG, (2) NED object name of group
  counterpart, (3) NED redshift of group counterpart, (4) Velocity
  difference between the SFCG and its group counterpart, (5) and (6)
  celestial coordinates of group counterpart, (7) separation distance
  on the sky between the SFCG and its group counterpart and (8)
  angular radius of the SFCG computed following Eq. \ref{eq:rtheta}.}
\label{tab:suspg}
\end{table}
\twocolumn

\begin{itemize}

\item SFCG092: Although, SDSSCGB~46521 is separated from this SFCG by
  $D_{\rm{}sep}$=5.8 arcmin, there is another closer ($D_{\rm{}sep}$=0.3
  arcmin) group (SDSSCGA~01937) with unknown redshift that may or may
  not be associated to it.

\item SFCG106: WBL 340 is its closer counterpart with $D_{\rm{}sep}$=4.8
  arcmin and it is clearly not physically linked to it. However, from
  the study presented in section \ref{ssec:big}, we know that this SFCG
  is embedded in a larger structure, ABELL~1318.

\item SFCG117: ABELL~1514 is separated from the SFCG117 by
  $D_{\rm{}sep}$=5.97 arcmin. The nearest group to SFCG117 is a
  four-member galaxy group SDSSCGB~62865 without known redshift
  ($D_{\rm{}sep}$=3.948 arcmin) which is also clearly separated from
  the SFCG.

\item SFCG134: $[$RPG97$]$ 236, this a three-member galaxy group with
  $D_{\rm{}sep}$=4.964~arcmin.

\item SFCG146: SDSSCGB\,49472, this SDSS group is clearly separated from
SFCG146 by a $D_{\rm{}sep}$=3.627~arcmin.

\item SFCG256: UGCl\,314, the spectroscopic redshift of this cluster
  was derived from just one galaxy redshift in projection
  ($D_{\rm{}sep}$=4.097~arcmin) of a previously identified Uppsala
  cluster.

\end{itemize}

In these examples, the matched groups are clearly not related to the
SFCG, therefore the 6 arcmin matching radius was somewhat too
liberal. From this fact together with the inspection of the SFCGs in
the GALEX and optical images, shown in the collection of stamps cited
at the end of section \ref{sec:gal_samp}, we conclude that a
crossmatching radius of 2.5 arcmin is an appropriate constraint to
link a newly identified SFCG with a previously known galaxy
group. \color{dreen} Applying this constraint we find 59 SFCGs with a
previously known group counterpart, 24 of which have known
spectroscopic redshifts.

\color{black}
\subsubsection{Search for group members without a bright UV counterpart}

We are also interested in finding other UV-faint galaxies with known
redshifts in the neighborhood of each SFCG, particularly for those
SFCGs with no redshift information. We derive a new additional
redshift estimation with the same prescription that we applied for
defining a redshift estimation for the set of UV galaxies belonging to
each SFCG (see Subsection \ref{ssec:gselec}), but now we include all
galaxies, UV-bright or not, within 2.5 arcmin from the SFCG
center. \color{dreen} The values for $\langle$$z_{\rm gal}$$\rangle$
correspond to the median redshift among those galaxies with known
redshifts inside a searching circle of $D_{\rm{}sep}$=2.5~arcmin. The
$g_{z}$ is the number of galaxies with known redshift within a
velocity interval of $\mid$$\Delta$v$_{\rm{}l-o-s}$$\mid$$\le$
10$^{3}$\,km\,s$^{-1}$ where $\Delta$v$_{\rm{}l-o-s}$ is the l-o-s
velocity difference with respect to the redshift $\langle$$z_{\rm
  gal}$$\rangle$. If there is no known redshift associated to the
galaxies of a specific SFCG, we label the redshift value with zero,
$\langle$$z_{\rm{}gal}$$\rangle$=0. The derived group redshifts
$\langle$$z_{\rm{}gal}$$\rangle$ are shown in Table \ref{tab:addz} and
the overall set of compiled galaxy redshifts is also overplotted in
the collection of stamps provided in Figure A1 of Appendix. This
approach yields 222 group redshifts $\langle$$z_{\rm gal}$$\rangle$
(79\% of the whole sample) with 87 SFCGs with $g_{z}$=1
($\approx$31\%) and 135 SFCGs with two or more redshifts (48\%). To
summarize, we provide three redshift estimations for the SFCGs:

\color{black}
\begin{itemize}

\item $z_{\rm{}UV}$ defined by the UV-bright galaxies of the SFCG in Table
\ref{tab:CS}, 

\item $z_{\rm{}NED}$ the NED redshift for a previously catalogued
  galaxy group in a neighbourhood of 2.5 arcmin in radius of the SFCG
  in Table \ref{tab:addz} and,

\item \textcolor{dreen}{$\langle$$z_{\rm{}gal}$$\rangle$ defined by
  the median of those galaxy redshifts found in a SFCG neighbourhood
  of 2.5 arcmin in radius in Table \ref{tab:addz}.}

\end{itemize}

\onecolumn
\begin{table}[!ht]
\caption{\bf Additional redshift counterparts}
\begin{center}
\resizebox{0.75\hsize}{!}{\bf 
\begin{tabular}{r l l l l l}
\hline
n$_{\rm{}ID}$&$\langle$$z_{\rm{}gal}$$\rangle$& $g_{z}$&$z_{\rm{}NED}$ & $D_{\rm{}sep}$ &  Group Name   \\
           &                              &       &             &     arcmin   &               \\  
     (1)   &   (2)                        &   (3) &    (4)      &  (5)         &     (6)       \\  
\hline      
    1   &     0          &       0   &     0          &      0       &   X                                     \\  
    2   &     0.017899   &       3   &     0.017800   &      1.497   &   HCG\_100                               \\  
    3   &     0.041719   &       1   &     0          &      0       &   X                                     \\  
    4   &     0          &       0   &     0          &      0       &   X                                     \\  
    5   &     0          &       0   &     0          &      0       &   X                                     \\  
    6   &     0          &       0   &     0          &      0       &   X                                     \\  
    7   &     0.093258   &       1   &     0          &      0       &   X                                     \\  
    8   &     0.013066   &       2   &     0          &      0       &   X                                     \\  
    9   &     0.036399   &       2   &     0          &      0       &   X                                     \\  
   10   &     0.051799   &       1   &     0          &      0       &   X                                     \\  
   11   &     0.093101   &       2   &     0          &      0       &   X                                     \\  
   12   &     0.019557   &       2   &     0          &      0       &   X                                     \\  
   13   &     0.041095   &       1   &     0          &      0       &   X                                     \\  
   14   &     0.077190   &       4   &     0.077000   &      0.120   &   SDSSCGA\_00909                         \\  
   15   &     0.063567   &       3   &     0.063600   &      1.409   &   ABELL\_2800                            \\  
   16   &     0.086353   &       1   &     0          &      0       &   X                                     \\  
   17   &     0          &       0   &     0          &      0       &   X                                     \\  
   18   &     0.159397   &       1   &     0          &      0       &   X                                     \\  
   19   &     0          &       0   &     0          &      0       &   X                                     \\  
   20   &     0.041299   &       2   &     0          &      0       &   X                                     \\  
 ...    &      ...       &      ...    &      ...       &     ...      &             ...                         \\  
  261   &     0.051225   &       4   &     0.050800   &      1.280   &   MCXC\_J1749.8+6823                     \\  
  262   &     0.046689   &       1   &     0          &      0       &   X                                     \\  
  263   &     0.055402   &       2   &     0          &      0       &   X                                     \\  
  264   &     0          &       0   &     0          &      0       &   X                                     \\  
  265   &     0          &       0   &     0          &      0       &   X                                     \\  
  266   &     0.031292   &       1   &     0          &      0       &   X                                     \\  
  267   &     0          &       0   &     0          &      0       &   X                                     \\  
  268   &     0          &       0   &     0          &      0       &   X                                     \\  
  269   &     0.070699   &       4   &     0          &      0       &   X                                     \\  
  270   &     0          &       0   &     0          &      0       &   X                                     \\  
  271   &     0.027636   &       2   &     0          &      0       &   X                                     \\  
  272   &     0.017462   &       2   &     0.015960   &      0.913   &   WBL\_178                               \\  
  273   &     0.069071   &       6   &     0          &      0       &   X                                     \\  
  274   &     0.045465   &       2   &     0          &      0       &   X                                     \\  
  275   &     0.085412   &       4   &     0.085000   &      0.129   &   SDSSCGA\_00370                         \\  
  276   &     0.067000   &       2   &     0          &      0       &   X                                     \\  
  277   &     0          &       0   &     0          &      0       &   X                                     \\  
  278   &     0.000907   &       1   &     0          &      0       &   SDSSCGB\_67914                         \\  
  279   &     0.063227   &       1   &     0.434700   &      0.579   &   WHL\_J130719.4+133902                  \\  
  280   &     0.039417   &       1   &     0          &      0       &   X                                     \\  
\hline
\end{tabular}}
\end{center}
{\bf (1) ID number of the SFCG, (2) median redshift for galaxies with
  known redshift given by NED inside $D_{\rm{}sep}$=2.5~arcmin, (3)
  number of galaxies to derive the redshift, (4) NED group redshift,
  (5) angular separation distance between the SFCG and the previously
  identified galaxy group and (6) preferred NED name of a previously
  identified group counterpart.}
\label{tab:addz}
\end{table}
\twocolumn

\subsection{Purity of the sample of SFCGs: Check of the projected and kinematical surroundings of SFCGs}
\label{ssec:velacc}

\textcolor{red}{Groups of galaxies selected in projection without the
  use of redshifts are subject to contamination by discordant-redshift
  galaxies.} \citet{Diaz-Gimenez&Mamon_2010} studied the purity of a
sample of compact groups selected with the Hickson's criteria from the
$z$=0 output of three semi-analytical models of galaxy formation run
on the Millennium Simulation. Among the ``complete´´ sample of compact
groups fulfilling the Hickson's criteria, roughly 60\% of their
different mock projected compact group samples have at least four
galaxies within 10$^{3}$\,km\,s$^{-1}$ of the median velocity of the
group.

Also, \citet{Mendel_et_al_2011} found that when compact groups are
selected according to the criteria of \citet{Hickson_1982}, roughly
half of the compact groups can be associated with relatively rich
structures, while the remaining half are likely to be either
independent structures in the field or associated with comparably poor
groups.

With the same aim of estimating the fraction of SFCGs that present a
minimum number of their galaxies with accordant l-o-s velocities, we
retrieve all those objects with spectra classified as `galaxy' in a
radius of 15 arcmin around each group from the 10th Data release of
the Main Galaxy Sample (MGS) of the Sloan Digital Sky Survey
(SDSS). There are 139 groups compiled with good SDSS/MGS
coverage. \textcolor{red}{Among the 321 UV group members from the SFCG
  sample with a SDSS/MGS counterpart, there are 308 objects (i.e.
  96\%) spectroscopically classified as galaxies
  ({\bf\,class}=`GALAXY'), 4 UV members (i.e. 1.2\%) classified as
  stellar objects ({\bf\,class}=`STAR') and 9 objects (i.e. 2.8\%)
  classified as quasars ({\bf\,class}=`QSO').} \color{dreen} The
radial and kinematical distributions of galaxies around the SFCGs
covered by the SDSS/MGS are added to the article as additional on-line
only material in Figures A2, A3 and A4 of Appendix.

\color{black} We classify the radial and kinematical distribution of
galaxies around each SFCG depending on three number counts:

\begin{itemize}

\item $n_{\rm{}c}$ number of galaxies inside a circle of 2.5 arcmin
  centered on the group without any l-o-s velocity constraint. We
  outline that this number corresponds just to those group galaxies
  covered by the Main Galaxy Sample of SDSS, knowing that total number
  of group galaxies should be larger due to the two situations
  described below.

\item $n_{\rm{}v}$ number of galaxies inside a circle of 2.5 arcmin
  centered on the group and within 1200\,km\,s$^{-1}$ centered at
  c$z_{\rm{}UV}$ or the l-o-s velocity of the galaxy nearest to the
  group center. This corresponds to the number of bound members of the
  SFCG observed in the SDSS/MGS.

\item $n_{\rm{}s}$ number of galaxies with a radial projected
  distance to the group center between 2.5 and 10 arcmin and within
  1200\,km\,s$^{-1}$ centered at c$z_{\rm{}UV}$ or the l-o-s velocity
  of the galaxy closest to the group center. This number is sampling
  the richness of a possible galaxy structure where the group is
  embedded.

\end{itemize}

\color{red} An SFCG is considered a real group when it fulfills either
of these two conditions:

\begin{itemize}

\item $n_{\rm{}v}$$\ge$3, i.e. at least three velocity-accordant and
  close galaxy members, or

\item $n_{\rm{}c}$$\ge$2 and $n_{\rm{}v}$$>$$n_{\rm{}c}$/2 i.e. more
  than half of galaxies observed by the SDSS/MGS being
  velocity-accordant and close galaxy members. This constraint
  produces that groups with $n_{\rm{}c}$=2 require having
  $n_{\rm{}v}$=2 and groups with $n_{\rm{}c}$=3 require having
  $n_{\rm{}v}$$\ge$2 to be considered as real groups.

\end{itemize}

Four situations describing the radial and kinematical distribution of
galaxies around each SFCG are considered:

\begin{itemize}

\item (a) a real group is considered isolated if
  $n_{\rm{}s}$$\le$5. This is not an extreme constraint of
  isolation. This constraint only tries to take into account that the
  UV members could belong to a group also containing other UV faint
  members and this group could be slightly extended beyond the 2.5
  arcmin radius. We are considering that a few galaxies surrounding
  the SFCG indicate a poor structure.

\item (b) a real group is considered to be embedded in a larger galaxy
  structure (e.g. cluster) if $n_{\rm{}s}$$\ge$6.

\item (c) the SFCG is considered partially the result of a projection
  effect if, for groups with $n_{\rm{}c}$$\ge$2 also
  $n_{\rm{}v}$$\le$$n_{\rm{}c}$/2 and $n_{\rm{}v}$$\le$2.

\item (d) the SFCG is framed in the low statistics case and
  consequently excluded from the previous situations when there is
  just one or no galaxies in the close neighbourhood of the SFCG i.e.
  $n_{\rm{}c}$$\le$1.

\end{itemize}

\color{black} The low statistics case can be the result of two
situations.  First, the known fiber collision problem in the Main
Galaxy Sample of SDSS; two spectroscopic fibers cannot be placed
closer than 55 arcsec on a given plate \citep{Strauss_et_al_2002}.
Knowing that the maximum linking length in the FoF algorithm is set to
1.5 arcmin, many of the group galaxies can be closer than the fiber
collision distance. In addition, galaxies in SFCGs, which are selected
in UV, can be too optically faint and the SDSS/MGS magnitude limit
$r'$$<$17.77 can reject these galaxies, or they can be relatively
bright and have not been observed due to the SDSS/MGS completeness
strongly decreasing below $r'$$\sim$14
\citep{Montero-Dorta&Prada_2009}. \color{blue} Specifically, for the
groups identified in a 14.5$\le$$r'$$\le$18 SDSS galaxy catalogue by
\citet{McConnachie_et_al_2009}, 43 per cent of galaxy members have
spectroscopic information available.

\color{black} For those 139 SFCGs observed by the SDSS/MGS, we obtain
the following distribution: 54 SFCGs are isolated groups, 8 SFCGs are
embedded in larger galaxy structures, 58 SFCGs are partially the
result from a projection effect and 19 SFCGs are considered
incompletely covered by the SDSS/MGS.  Then, in the case of the group
subsample with enough statistics
(a)+(b)+(c)=\allowbreak{}54+8+58=\allowbreak{}120, we see that a
fraction of 52\% ((a)+(b)=62 SFCGs) have a significant number of
galaxies with accordant l-o-s velocities. We thus expect that roughly
half of the entire SFCG sample will be caused by chance
projections. \color{red} It is also found that the fraction of real
groups $f_{\rm{}rg}$ does not present a monotonic trend with
$n_{\rm{}c}$ (i.e. the number of galaxies with known redshift in the
close neighborhood of the group): \color{dreen}
$f_{\rm{}rg}$=47$\pm$11\% for $n_{\rm{}c}$=2,
$f_{\rm{}rg}$=62.5$\pm$8.6\% for $n_{\rm{}c}$=3 and
$f_{\rm{}rg}$=47.8$\pm$6.0\% for $n_{\rm{}c}$$>$3.

\color{blue} With respect to the close environment of groups, we find
a high fraction of isolated groups 54/(54+8)=87\%. This relatively
high fraction comes from the fact that we are defining here an
isolated group as that group which simply is not surrounded by a large
galaxy structure such a galaxy cluster but they are clearly
distinguishable from its surroundings.

\color{black} 
\section{Conclusions}

This article provides a local sample ($z$$\lesssim$0.15) of compact
groups of star-forming galaxies.  In this type of groups, galaxies
strongly interact among themselves and with the rest of the group
components (ICM, dark matter halo). This induces morphological changes
and star formation events which are currently taking place. The
peculiar evolutionary stage of these groups provides a wealth of
galaxy observables that may clarify the theoretical framework about
galaxy evolution in groups.

We have performed an all-sky search for compact groups of star-forming
galaxies in the GALEX UV catalogues. \color{red} Starting from the
1447 groups initially identified by the Friends-of-Friends group
finder, 960 were identified as groups of stellar and/or galactic
objects at the outskirts of the two Magellanic Clouds, while another
about 200 groups were removed because they did not have, at least 3 UV
members compiled as galaxies by NED or, two galaxies within
$\Delta$$z$/(1+$z$)=0.004. \color{black} The result from this search
are 280 galaxy groups composed by 226, 39, 11 and 4 groups with four,
five, six and seven UV members, respectively. Only 59 of the 280
identified SFCGs present a previous catalogued group counterpart in
their neighbourhood $D_{\rm{}sep}$$<$2.5 arcmin. Those groups with a
good SDSS spectroscopic coverage show an important fraction of them,
roughly half, having a significant number of galaxies with accordant
l-o-s velocities. At least, 26 SFCGs are embedded in the infall region
of previously catalogued galaxy clusters. A compilation of galaxy
redshifts in a sky region of 2.5 arcmin around the SFCG center
provides group redshifts for a total of 222 group candidates.

\color{red} As in all group catalogues, accurate redshifts are
required on all galaxy members to remove interlopers or identify
blended galaxy objects. In this respect, a full redshift survey of
this sample is underway as well as an H$\alpha$ survey of the groups
with known redshift. These results may elucidate several open issues
about the evolution of compact groups. \color{black}

\section*{Acknowledgements}
\label{sec:acknowledgements}

The authors would like to thank the referee, Dr. Gary Mamon, for the
substantial improvements on the original manuscript that resulted from
his recommendations. The article have been improved also thanks to the
helpful comments of Christopher P. Haines. J.D.H.F. acknowledges
support through the FAPESP grant project
2012/13381-0. C.M.d.O. acknowledges support through FAPESP project
2006/56213-9 and CNPq grant 305205/2010-2. The formal acknowledgements
to the resources used in this work can be read at:
www.sdss.org\allowbreak{}/dr6\allowbreak{}/coverage\allowbreak{}/credits.html
for the Sloan Survey and ned.ipac.caltech.edu for the NED webpage and
galex.stsci.edu\allowbreak{}/GR6\allowbreak{}/?page=acknowledgments
for the GALEX mission. The authors have used the TOPCAT software
\citep{Taylor_2005} in this work.


\end{document}